\documentclass[8pt]{article}
\RequirePackage[OT1]{fontenc}
\usepackage{natbib}
\RequirePackage[colorlinks,citecolor=blue,urlcolor=blue]{hyperref}
\usepackage[english]{babel}
\usepackage{amsthm}
\usepackage{amsmath}
\usepackage{amsfonts}
\usepackage{amssymb}
\usepackage{graphicx}
\usepackage{enumerate}
\usepackage{color}
\usepackage{array}
\usepackage{tikz}
\usepackage{comment}
\usepackage{bm}
\usepackage{titlesec}
\usepackage{float}
\usepackage{multirow}
\usepackage{babel}
\usepackage[font=small,labelfont=bf]{caption}
\usepackage[utf8]{inputenc}
\usepackage[toc,page]{appendix}
 \usepackage{bbm}

\allowdisplaybreaks
\usepackage[algoruled,boxed,lined]{algorithm2e} 
\makeatletter 
\g@addto@macro{\@algocf@init}{\SetKwInOut{Parameter}{Parameters}} 
\makeatother
 \usepackage{mathtools}
 
 \usepackage{verbatim}

\definecolor{myorange}{HTML}{E24100}
\definecolor{mygreen}{HTML}{228b22}

\def\frac#1#2{{\textstyle{#1\over#2}}}

\DeclareSymbolFont{AMSb}{U}{msb}{m}{n}
\DeclareMathSymbol{\Natural}{\mathbin}{AMSb}{"4E}
\DeclareMathSymbol{\Integer}{\mathbin}{AMSb}{"5A}
\DeclareMathSymbol{\Real}{\mathbin}{AMSb}{"52}
\DeclareMathSymbol{\Rational}{\mathbin}{AMSb}{"51}
\DeclareMathSymbol{\Imaginary}{\mathbin}{AMSb}{"49}
\DeclareMathSymbol{\Complex}{\mathbin}{AMSb}{"43} 
\DeclareMathSymbol{\Disk}{\mathbin}{AMSb}{"44} 
\def\bi{\begin{itemize}}
\def\ei{\end{itemize}}
\def\bd{\begin{description}}
\def\ed{\end{description}}
\def\ben{\begin{enumerate}}
\def\een{\end{enumerate}}
\def\bv{\begin{verbatim}}
\def\ev{\end{verbatim}}

\def\hat#1{{\widehat{#1}}}

\newcommand{\bs}{\boldsymbol}

\def\pr{{\rm Pr}}
\def\Pr{\pr}

\def\2to{{\ {\buildrel 2\over \longrightarrow}\ }}

\def\I1ton{{$I_1,\ldots,I_n$}}
\def\X1ton{{$X_1,\ldots,X_n$}}
\def\Y1ton{{$Y_1,\ldots,Y_n$}}
\def\Z1ton{{$Z_1,\ldots,Z_n$}}
\def\R1ton{{$R_1,\ldots,R_n$}}
\def\e1ton{{$e_1,\ldots,e_n$}}
\def\t1ton{{$t_1,\ldots,t_n$}}
\def\x1ton{{$x_1,\ldots,x_n$}}
\def\y1ton{{$y_1,\ldots,y_n$}}
\def\z1ton{{$z_1,\ldots,z_n$}}

\newtheorem{defn}{Definition}

\newtheorem{cor}[defn]{Corollary}
\newtheorem{prop}[defn]{Proposition}

\usepackage[margin=1in]{geometry}

\begin{document}
\thispagestyle{empty}
\baselineskip=28pt
\vskip 5mm
\begin{center} 
{\Large{\bf Partial Tail-Correlation Coefficient \\Applied to Extremal-Network Learning}}
\end{center}
\baselineskip=12pt
\vskip 5mm

\begin{center}
\large 
Yan Gong$^1$, Peng Zhong$^1$, Thomas Opitz$^2$, Rapha\"el Huser$^1$
\end{center}

\footnotetext[1]{
\baselineskip=10pt Statistics Program, Computer, Electrical and Mathematical Sciences and Engineering (CEMSE) Division, King Abdullah University of Science and Technology (KAUST), Thuwal 23955, Saudi Arabia. E-mails: yan.gong@kaust.edu.sa; peng.zhong@kaust.edu.sa; raphael.huser@kaust.edu.sa.}
\footnotetext[2]{
\baselineskip=10pt Biostatistics and Spatial Processes, INRAE, Avignon 84914, France. E-mail: thomas.opitz@inrae.fr.}

\baselineskip=17pt
\vskip 4mm
\centerline{\today}
\vskip 6mm

\begin{center}
{\large{\bf Abstract}}
\end{center}
We propose a novel extremal dependence measure called the partial tail-correlation coefficient (PTCC), in analogy to the partial correlation coefficient in classical multivariate analysis. The construction of our new coefficient is based on the framework of multivariate regular variation and transformed-linear algebra operations. We show how this coefficient allows identifying pairs of variables that have partially uncorrelated tails given the other variables in a random vector. Unlike other recently introduced conditional independence frameworks for extremes, our approach requires minimal modeling assumptions and can thus be used in exploratory analyses to learn the structure of extremal graphical models. Similarly to traditional Gaussian graphical models where edges correspond to the non-zero entries of the precision matrix, we can exploit classical inference methods for high-dimensional data, such as the graphical LASSO with Laplacian spectral constraints, to efficiently learn the extremal network structure via the PTCC. We apply our new method to study extreme risk networks in two different datasets (extreme river discharges and historical global currency exchange data) and show that we can extract meaningful extremal structures with meaningful domain-specific interpretations.

\baselineskip=16pt

\par\vfill\noindent
{\bf Keywords:} Extreme event; Graphical Lasso; Multivariate regular variation; Network structure learning; Partial tail dependence.\\

\pagenumbering{arabic}
\baselineskip=26pt

\newpage

\section{Introduction}\label{sec:introduction}
Characterizing the extremal dependence of complex stochastic processes (e.g., in spatial, temporal, and spatio-temporal settings)  is fundamental for both statistical modeling and applications, such as risk assessment in environmental and financial contexts. Important applications include the modeling of precipitation extremes \citep{Huser.Davison:2014a,opitz2018inla,bacro2020hierarchical,saunders2021regionalisation,richards2022modelling}, heatwaves \citep{winter2016modelling,zhong2022modeling}, and air pollution \citep{Vettori.etal:2019, Vettori.etal:2020}, as well as financial risk assessment \citep{bassamboo2008portfolio,ferro2011extremal,marcon2016bayesian,bekiros2017extreme,yan2019cross,gong2019asymmetric}. Often, applications illustrate the benefits of methodological innovations, such as the application of the extremal dependence measure in \citet{larsson2012extremal}, which is a key ingredient of the present work, in the analysis of financial data. 

Models for extremal dependence traditionally rely on asymptotic frameworks, such as  max-stable processes for block maxima  or $r$-Pareto processes for threshold exceedances of a summary functional of the process over a high threshold. Recently, more advanced models have been proposed to further improve flexibility, especially towards modeling of asymptotically independent data with dependence vanishing at the most extreme levels, such as inverted max-stable processes \citep{wadsworth2012dependence}, max-mixture models \citep{ahmed2020spatial}, random scale mixtures of Gaussian processes \citep{Opitz2016,Huser2017} or of more general processes \citep{wadsworth2017modelling,engelke2019extremal,huser2019modeling}, max-infinitely divisible processes \citep{bopp2021hierarchical}, and conditional spatial extremes models \citep{wadsworth2019higher}; for a comprehensive review, see \citet{huser2022advances}. Specifics of serial extremal dependence have been studied by \citet{davis2013measures}, among others. 

In the study of stochastic dependence structures, networks and graphs are natural tools to represent dependence relationships in  multivariate data. Conditional independence, sparsity, and parsimonious representations are key concepts in graph-based approaches for random vectors. 
Recently, graph-based tools have also been  developed for extremal dependence, where variants of conditional independence apply to variables not directly connected by the edges of the graph. For example, \citet{huang2019new} provide an exploratory tool, called the \emph{$\chi\text{-}$network,} for modeling extremal dependence, and they use it to analyze maximum precipitation during the hurricane season in the United States (US) Gulf Coast and in surrounding areas. In their approach, however, the $\chi$-network does not remove the effect of confounding variables, so it does not naturally lead to sparse extremal dependence representations. More recently, \citet{engelke2020graphical} introduce a  notion of conditional independence adapted to multivariate Pareto distributions arising for limiting multivariate threshold exceedances, and they use it to develop {parametric graphical models for extremes based on the H\"usler--Reiss distribution}. Similarly, \citet{gissibl2018graphical} and \citet{kluppelberg2021estimating} propose max-linear constructions for modeling maxima on tree-like supports, and \citet{tran2021estimating} propose {QTree}, a simple and efficient algorithm to solve the ``Latent River Problem" for the important case of extremes on trees. In the same vein, \citet{engelke2020structure} develop a data-driven methodology for learning the graphical structure in the setting of \citet{engelke2020graphical}, whereas \citet{rottger2021total} further propose H\"usler--Reiss graphical models under the assumption of multivariate total positivity of order two (MTP$_2$), which allows estimating sparse graphical structures. Finally,  \citet{engelke2021sparse} review the recent developments in sparse representations, dimension reduction approaches, and graphical models for extremes. Overall, existing graphical representations for extremes from the literature often rely on rather stringent asymptotically justified models, sometimes leading to issues when dealing with relatively high-dimensional problems or when specific graph structure assumptions (e.g., trees) are required. A recent exception is \citet{Engelke.etal:2022}, who develop theories and methods for learning extremal graphical structures in high dimensions based on $L_1$ regularized optimization, though their methodology still assumes a parametric extremal dependence structure of H\"usler--Reiss type.


By contrast, rather than restricting ourselves to a strict parametric modeling framework, we adopt a more pragmatic and empirical approach. Specifically, our goal is to extend and enrich existing approaches by defining the new concept of \emph{partial tail correlation} as an extreme-value analog of the notion of partial correlation widely used in classical multivariate analysis, and by introducing a new coefficient that enables estimation of general extremal networks under minimal modeling assumptions. In the same way that correlation does not imply independence in general, our concept of \emph{partial tail-uncorrelatedness} is a weaker assumption than \emph{conditional tail independence}. However, we shall show that it still provides relevant insights into various forms of extremal dependence structures and helps in guiding modeling choices at a data exploratory stage.

As a novel extremal dependence measure, we propose the \emph{partial tail correlation coefficient (PTCC)} as an equivalent of the partial correlation coefficient in the non-extreme setting. In the classical setting,  the Pearson correlation coefficient between two random variables can give misleading interpretations when there are confounding variables that influence both variables, whereas the partial correlation coefficient  measures the residual degree of association between two  variables after the linear effects of a set of other variables have been removed. To compute the partial correlation between two variables of interest, we regress each of these variables  onto the set of covariates given by all the other variables in the multivariate random vector, and then compute the correlation between the residuals from the two fitted linear regressions. In the Gaussian setting, a partial correlation of zero is equivalent to conditional independence between two variables \citep{lawrance1976conditional, baba2004partial}, and the elements of the inverse of the covariance matrix (i.e., the \emph{precision matrix}) of the full vector are known to characterize this conditional (in)dependence structure. In this paper, we adopt a similar strategy to define the PTCC, namely by computing a suitable measure of tail dependence between residuals obtained by regressing variables using transformed-linear operations that do not alter tail properties. While classical linear regression only makes sense for Gaussian-like data, such transformed-linear operations can be used for tail regression of multivariate regularly-varying random vectors, which is a fundamental assumption characterizing asymptotically dependent extremes.

To be more precise, we here define the PTCC by building upon the framework of \citet{cooley2019decompositions}, who developed a customized transformed-linear algebra on the positive orthant, preserving multivariate regular variation and thus being well adapted to ``linear'' methods for joint extremes. \citet{cooley2019decompositions} used this framework for principal component analysis of extremes based on decompositions of the so-called tail pairwise dependence matrix (TPDM), which conveniently summarizes information about extremal dependence in random vectors and possesses appealing properties for such decompositions. The TPDM can be thought of as an analogy of the classical covariance matrix but tailored for multivariate extremes. In some follow-up work, \citet{mhatre2020transformed} then developed non-negative regularly varying  time series models with autoregressive moving average (ARMA) structure using the transformed-linear operations for time series extremes. For spatial extremes, \citet{fix2021simultaneous} extended the simultaneous autoregressive (SAR) model under the transformed-linear framework and developed an estimation method to minimize the discrepancy between the TPDM of the fitted model and an empirically estimated TPDM. Furthermore, \citet{lee2021transformed} recently introduced transformed-linear prediction methods for extremes. In the aforementioned papers, the TPDM always plays a central role. Similar to covariances, the entries of the TPDM are tail dependence measures giving insights into the direct extremal dependence structure without removing the influence of other confounding variables. However, just as the covariance matrix does not reflect partial correlations, the TPDM does not directly inform us about  partial associations among extremes. In this work, we fill this gap with our new proposed PTCC. Thanks to its definition in terms of transformed-linear operations, we show that the PTCC inherits several appealing features of the classical partial correlation coefficient. In particular, the PTCC between two components $X_i$ and $X_k$ from a random vector $\bm X$ is such that the $(i,k)$th entry of the inverse TPDM matrix of $\bm X$ equals zero if and only if the corresponding PTCC for these two variables is also equal to zero. In other words, \emph{partial tail-uncorrelatedness} can be conveniently read off from the zero elements of the inverse TPDM, similar to classical Gaussian graphical models. We then exploit this property to define a new class of extremal graphical models based on the PTCC and then use efficient inference methods to learn the extremal network structure from high-dimensional data based on state-of-the-art techniques from graph theory (e.g., the graphical LASSO with or without Laplacian spectral constraints). We note that here, our focus is on studying undirected graph structures, which is different from causal inference, where causal relationships can be encoded using directed graph edges. 

The remainder of this article is organized as follows. In Section~\ref{sec:ptcc}, we first review the necessary background on multivariate regular variation and transformed-linear algebra, as introduced in \citet{cooley2019decompositions}. Then, we define the new PTCC and the related notion of partial tail-uncorrelatedness. 
In Section~\ref{sec:extnet}, we present methods for learning general extremal network structures from the PTCC in a high-dimensional data setting, and we discuss two particularly appealing approaches, namely the graphical LASSO and Laplacian spectral constraint-based methods. Section~\ref{sec:simulation} presents a simulation study for general structured undirected graphs using the above two inference methods. In Section~\ref{sec:applications}, we apply these new tools to explore the risk networks formed by river discharges observed at a collection of monitoring stations in the upper Danube basin, and by historical global currency exchange rate data from different historical periods, covering different economic cycles, the COVID-19 pandemic, and the 2022 military conflict in Ukraine.





\section{Transformed-linear algebra for multivariate extremes}\label{sec:ptcc}
Before introducing the partial tail correlation coefficient (PTCC) and the related notion of partial-tail uncorrelatedness, we first briefly review the multivariate regular variation framework, which is our main assumption for defining the PTCC, and we also summarize the foundations of transformed-linear algebra.
\subsection{Regular variation framework and transformed linear algebra}\label{sec:regvar}
A random vector is multivariate regularly varying \citep{resnick2007heavy} (i.e., jointly heavy-tailed) if its joint tail decays like a power function. Precisely, we say that a $p$-dimensional random vector $\bs X\in \mathbb{R}_+^p = [0,\infty)^p$ with $p\in\mathbb{N}$ is \emph{regularly varying} if there exists a sequence $b_n \to\infty$ such that
\begin{equation}\label{eq:mrv}
n\,\Pr(b_n^{-1} \bs X\in\cdot) \xrightarrow{v} \nu_{\bs X}(\cdot),\quad n\to \infty,
\end{equation}
where $\xrightarrow{v}$ denotes vague convergence to the non-null limit measure $\nu_{\bs X}$, a Radon measure defined on the space  $[0,\infty]^p\setminus \{0\}$. This measure has the scaling property $r^\alpha\nu_{\bs X}(rB)=\nu_{\bs X}(B)$ for $r>0$ and Borel sets $B \subset [0,\infty]^p\setminus \{0\}$, where $\alpha>0$ controls the tail decay (with $1/\alpha$ commonly called the tail index).   For this reason, the measure can be further decomposed into a radial measure and an angular measure $H_{\bs X}$ on the unit sphere $\mathbb{S}_{p-1}^+ = \{\bs w \in \mathbb{R}_+^p: ||\bs w||_2 = 1\}$, such that $\nu_{\bs X}(\{\bs x \in [0,\infty]^p\setminus \{0\}: \|\bs x\|_2 \geq r, \ \bs x/\|\bs x\|_2 \in B_H)=r^{-\alpha}\times H_{\bs X}(B_H)$ for $r>0$ and Borel subsets $B_H$ of $\mathbb{S}_{p-1}^+$. The normalizing sequence $b_n$ is not uniquely determined but must satisfy $b_n = L(n)n^{1/\alpha}$, where $L(n)$ is a slowly varying function (at infinity), i.e., $L(n)>0$ and  $L(rn)/L(n)\rightarrow 1$ for any $r>0$, as $n\rightarrow\infty$.  We use the short-hand notation $\bs X \in$ RV$_+^{p}(\alpha)$ for a regularly varying vector $\bs X$ with tail index $1/\alpha$. 

\citet{cooley2019decompositions} introduced the \emph{transformed-linear algebra} framework to construct an inner product space on an open set (the so-called \emph{target space}) via a suitable transformation, where the distribution of the random vector $\bm X$ has support within this set. Our use will mainly concern transformation towards the target space $\mathbb{R}_+^p$ from the space $\mathbb{R}^p$, but we first present the general approach. Let $t$ be a bijective transformation from $\mathbb{R}$ onto some open set $\mathbb{X}\subset\mathbb{R}$, and let $t^{-1}$ be its inverse. For a $p$-dimensional vector $\bs y \in \mathbb{R}^p$, we define $\bs x = t(\bs y)\in \mathbb{X}^p$ componentwise. Then, arithmetic operations among elements of the target space are carried out in the space $\mathbb{R}^p$ before transforming back to the target space. We define vector addition in $\mathbb{X}^p$ as $\bs x_1\oplus \bs x_2 = t\{t^{-1}(\bs x_1) + t^{-1}(\bs x_2)\}$, and scalar multiplication with a factor $a\in\mathbb{R}$  as $a\circ \bs x = t\{at^{-1}(\bs x)\}.$ The additive identity in $\mathbb{X}^p$ is set to $\bs 0_{\mathbb{X}^p}=t(\bs 0)$, and the additive inverse of $\bs x\in\mathbb{X}^p$ is given as $\ominus \bs x=t\{-t^{-1}(\bs x)\}$. A valid inner product between two elements $\bs x_1=(x_{1,1},\ldots,x_{1,p})^T,\bs x_2=(x_{2,1},\ldots,x_{2,p})^T\in\mathbb{X}^p$ from the target space is then obtained by applying the usual scalar product in $\mathbb{R}^p$, i.e., we set $\langle \bs x_1,\bs x_2\rangle =\sum_{j=1}^p t^{-1}(x_{1,j})t^{-1}(x_{2,j})$. To obtain an inner product space on the positive orthant for which arithmetic operations preserve multivariate regular variation, thus having a negligible effect on large values, we follow \citet{cooley2019decompositions} and define the specific transformation $t: \mathbb{R}\mapsto (0, \infty)$ given by
$$
t(y) = \log\{1+\exp(y)\},
$$
though they are other possibilities. We have $y/t(y)\rightarrow 1$ as $y\rightarrow \infty$, such that the upper tail behavior of a random vector $\bs Y=t(\bs X)$ is preserved through $t$. For lower tails, we have $\exp(y)/t(y)\rightarrow 1$ as $y\rightarrow -\infty$. The inverse transformation is $t^{-1}(x) = \log(\exp(x)-1)$, $x>0$. Algebraic operations done in the vector space induced by the above transformation $t$ are commonly called \emph{transformed-linear operations}, and we can exploit this framework to extend classical linear algebra methods (e.g., principal component analysis, etc.) to the multivariate extremes setting, where heavy-tailed vectors and models are often conveniently expressed on the positive orthant, $\mathbb R^p_+$. We note that our main assumption, multivariate regular variation, implies asymptotic tail dependence (as well as homogeneity of the limit measure in \eqref{eq:mrv}), but it does not impose further parametric structural assumptions such as with Pareto models of H\"usler--Reiss type.  

\subsection{Inner product space of regularly varying random variables}

With transformed-linear operations, we can use a vector of independent and identically distributed (i.i.d.) random variables to construct new regularly varying random vectors on the positive orthant that possess tail dependence. Suppose that $\bs Z = (Z_1,\ldots,Z_q)^T \geq 0$ is a vector of $q\in\mathbb{N}$ i.i.d.\ regularly varying random variables with tail index $1/\alpha$, such that there exists a sequence $\{b_n\}$ that yields
$$
n\,\Pr(Z_j>b_n z)\to z^{-\alpha}, \quad n\,\Pr\{Z_j\leq \exp(-k b_n)\}\to 0, \quad k>0, \quad j=1,\ldots,q,
$$
where the first condition is equivalent to regular variation  \eqref{eq:mrv} in dimension $p=1$.
The random vector $\bm Z$ with independent components has a limit measure of multivariate regular variation characterized by $\nu_{\bm Z}\left\{[\bm 0, \bm z]^C\right\}=\sum_{j=1}^q z_j^{-\alpha}$ for $\bm z=(z_1,\ldots,z_q)^T > \bm 0$. 
Then, we can construct new regularly varying $p$-dimensional random vectors $\bs X = (X_1, \ldots,X_p)^T$ by exploiting transformed-linear operations, via a matrix product with a deterministic matrix $A = (\bs a_1, \ldots, \bs a_q) \in  \mathbb{R}_+^{p\times q},$ with columns $\bs a_j\in  \mathbb{R}_+^p$, as follows:
\begin{equation}\label{eq:transformed-linear construction}
    \bs X = \overset{q}{\underset{j=1}{\oplus}}\bs a_j\circ Z_j.
\end{equation}
We write $\bs X = A \circ \bs Z \in$ RV$_+^{p}(\alpha)$. This construction ensures that the multivariate regular variation property is preserved with the same index $\alpha$  \citep[Corollary 1,][]{cooley2019decompositions}. Furthermore, we require $A$ to have a full row-rank. Based on the construction  \eqref{eq:transformed-linear construction}, it is possible to define a (different) inner product space spanned by  the random variables obtained by transformed-linear operations on $\bs Z$, where some but not all  of the components of $\bs a_j$ are further allowed to be non-positive. 
Following  \citet{lee2021transformed}, an inner product of $\langle X_i,X_k\rangle $ on the space spanned by all possible transformed-linear combinations of the elements of the random vector $\bs X$ constructed as in \eqref{eq:transformed-linear construction}, may be defined as follows:
 \[\langle X_i, X_k\rangle  = \sum_{j=1}^q a_{ij} a_{kj},\]
 where $a_{ij}$ refers to the entry in row $i$ of the column $j$ of the matrix $A$ for $i\in\{1,\ldots,p\}$ and $j\in\{1,\ldots,q\}$,
and the corresponding norm becomes $||X|| = \sqrt{\langle X, X\rangle }$. The metric induced by the inner product is $d(X_i, X_k) = ||X_i\ominus X_k||=[\sum_{j=1}^q(a_{ij}-a_{kj})^2]^{1/2},$ for $i,k = 1,\ldots, p.$

\subsection{Generality of the framework}

In practice, given a random vector $\bm X$ for which we assume $\bs X \in$ RV$_+^{p}(\alpha)$, we will further assume that it allows for a stochastic representation as in \eqref{eq:transformed-linear construction}. Since the constructions of type \eqref{eq:transformed-linear construction} form a dense subclass of the class of multivariate regularly varying vectors  (if $q$ is not fixed but allowed to tend to infinity, i.e., $q\rightarrow\infty$), this assumption is not restrictive; see \citet{fougeres2013dense} and \citet{cooley2019decompositions}. 

Thanks to their flexibility, representations akin to the transformed-linear random vectors in \eqref{eq:transformed-linear construction} have recently found widespread interest in statistical learning for extremes. The fundamental model structure used in the causal discovery framework for extremes developed by \citet{gnecco2021} is essentially based on a variant of  \eqref{eq:transformed-linear construction}. 
In the setting of max-linear models, in particular the graphical models of \citet{gissibl2018graphical},  we can use \eqref{eq:transformed-linear construction} to construct random vectors $\bs X$ possessing the same limit measure $\nu_{\bs X}$ as the max-linear vectors. Finally, low-dimensional representations of extremal dependence in random vectors obtained through variants of the $k$-means algorithm can be shown to be equivalent to the extremal dependence induced by construction \eqref{eq:transformed-linear construction}; see \citet{janssen2020}.  
\subsection{Tail pairwise dependence matrix} 

The tail pairwise dependence matrix \citep[TPDM,][]{cooley2019decompositions} is defined to summarize the pairwise extremal dependence of a regularly varying random vector using the second-order properties of its angular measure. Let $\alpha = 2$, which ensures desirable properties; in practice, this condition can be ensured through appropriate marginal pre-transformation of data \citep{cooley2019decompositions}. Then, the TPDM $\Sigma$ of $\bs X\in$ RV$_+^{p}(2)$ is defined as follows:
$$
\Sigma=(\sigma_{ik})_{i,k=1,\ldots,p},\qquad \text{with}\qquad \sigma_{{ik}} := \int_{\mathbb{S}_{p-1}^+}w_i w_k\text{d}H_{\bs X}(\bs w),
$$
where $H_{\bs X}$ is the angular measure on $\mathbb{S}_{p-1}^+ = \{\bs w \in \mathbb{R}_+^p: ||\bs w||_2 = 1\}$ as introduced in Section~\ref{sec:regvar}. The matrix $\Sigma$ is an extreme-value analog of  the covariance matrix, and it has similar useful properties. It is positive semi-definite and completely positive, i.e., there exists a finite $p\times q$ matrix $A$ with nonnnegative entries such that the TPDM can be factorized as $\Sigma=AA^T$ \citep[][Proposition~5]{cooley2019decompositions}. The matrix $A$ is not unique, in particular if we do not impose nonnegative entries.
Specifically, for random vectors $\bm X$ obtained by the transformed-linear construction \eqref{eq:transformed-linear construction}, the entries of $\Sigma$ correspond to the values of the inner product $\sigma_{ik}=\langle X_i,X_k\rangle $. 
In the following, we further assume that $\Sigma$ is positive definite, which guarantees the existence of the inverse matrix of the TPDM. 

We emphasize that the special case where $\sigma_{ik}=0$ is equivalent to  asymptotic tail independence of the components $X_i$ and $X_k$ \citep[see][]{cooley2019decompositions}, meaning that the conditional exceedance probability $\mathrm{Pr}(F_{X_i}(X_i)>u \mid F_{X_k}(X_k)>u)$ tends to $0$ as $u\to1$ \citep{sibuya1960bivariate, ledford1996statistics}.  

By exploiting the property that the TPDM is completely positive, we can construct new transformed-linear random vectors that have the same TPDM as a given random vector $\bs X$. Since we can always factorize the TPDM as $\Sigma=AA^T$ for some matrix $A$ of dimension $p\times q$, we can then use the construction  \eqref{eq:transformed-linear construction} by multiplying $A$ with a (new) random vector $\bs Z\in \text{RV}_+^{p}(2)$ of independent regularly varying random variables, and the resulting vector will still have TPDM $\Sigma$. It is also worth noting that as $q\to\infty$, the angular measure of the new random vector can be arbitrarily close to that of $\bs X$ thanks to the denseness property of discrete angular measures. While $A$ is not unique, we note that it is very important that the inner product depends only on the entries of the matrix $\Sigma = AA^T$, such that the specific choice of $A$ does not matter. 
 
An estimator of the TPDM was proposed by \citet[][Section~7.1]{cooley2019decompositions}. For an i.i.d.\ sequence of vectors $\bs x_t$, $t=1,\ldots,n_{\text{samp}}$, i.e., independent realizations from a random vector $\bs X\in$ RV$_+^{p}(2)$, define 
 \begin{equation}\label{eq:estimator}
\hat{\sigma}_{{ik}} = \hat{m}\int_{\mathbb{S}_{p-1}^+}w_i w_k\text{d}\hat{N}_{\bs X}(w)=\hat{m}\, n^{-1}_{\text{ext}}\sum^{n_{\text{samp}}}_{t=1}w_{ti}w_{tk}\mathbbm{1}(r_t>r_0),
 \end{equation}
where $r_t = ||\bs x_t||_2$, $w_t=\bs x_t/r_t$, $r_0$ is a high threshold for the radial component, $n_{\text{ext}}=\sum_{t=1}^{n_{\text{samp}}}\mathbbm{1}(r_t>r_0)$ refers to the number of threshold exceedances, and the probability measure $N_{\bs X}(\cdot)=m^{-1}H_{\bs X}(\cdot)$ (with $m = H_{\bs X}(\mathbb{S}_{p-1}^+)$) is obtained by normalizing $H_{\bs X}$. Moreover, $\hat{m}$ denotes an estimate of $H_{\bs X}(\mathbb{S}_{p-1}^+)$, and $\hat{N}_{\bs X}$ is the empirical counterpart of $N_{\bs X}$. We note that when the data are preprocessed to have a common unit scale, we can set $m=p$ and there is no need to estimate the normalizing factor. The estimator \eqref{eq:estimator} was discussed by \citet{larsson2012extremal} in the bivariate case.
\subsection{Partial tail correlation coefficient}

In this section, we introduce our new measure, the partial tail correlation coefficient (PTCC), which is analogous to the partial correlation coefficient but tailored to heavy-tailed random vectors. 

Let $\bs X = A \circ \bs Z \in$ RV$_+^{p}(\alpha)$ be a $p$-dimensional vector constructed as in \eqref{eq:transformed-linear construction}, with TPDM $\Sigma$. We write $\bs X_{ik}=(X_i,X_k)^T$; $\bs X_{{\rm rest}}$ for the $(p-2)$-dimensional random vector obtained by removing the two components $X_i$ and $X_k$ from $\bs X$; $A_{ik}$ for the matrix comprising the $i$-th and $k$-th columns of $A$; and $A_{{\rm rest}}$ for the matrix $A$ without its $i$-th and $k$-th columns. Moreover, we define the $p$-dimensional random vector $\bs X'$ by re-ordering the columns of $\bs X$ as 
$\bs X' = (\bs X_{ik}^T, \bs X_{{\rm rest}}^T)^T= (A_{ik}, A_{{\rm rest}})\circ \bs Z.$ It is straightforward to show that the best transformed-linear predictor of $\bs X_{ik}$ given $\bs X_{{\rm rest}}$ can be obtained as $\hat{\bs X}_{ik}=\bs B\circ\bs X_{{\rm rest}} = (\hat{X}_{ik}^i, \hat{ X}_{ik}^k)^T$, where $\bs B = (\bs b_1, \bs b_2)^T$ is a $2\times(p-2)$ matrix with $\bs b_1, \bs b_2\in\mathbb R_+^{p-2}$ chosen such that $d(X_{i}, \hat{X}_{ik}^i)$ and $d(X_{k}, \hat{X}_{ik}^k)$ attain their minimum, respectively. Suppose that the TPDM of $\bs X'$ is of the block-matrix form
$$
\Sigma_{\bs X'} = 
\begin{bmatrix}
\Sigma_{ik,ik} & \Sigma_{ik,{\rm rest}}\\
\Sigma_{{\rm rest},ik} & \Sigma_{{\rm rest},{\rm rest}}	
\end{bmatrix}
$$
where $\Sigma_{ik,ik}\in\mathbb R_+^{2\times 2}$ is the TPDM restricted to $\bs X_{ik}$, i.e., $\Sigma_{ik,ik}=\text{TPDM}(\bs X_{ik})=\begin{bmatrix}
\Sigma_{ii} & \Sigma_{ik} \\
\Sigma_{ki} & \Sigma_{kk}
\end{bmatrix}$; $\Sigma_{{\rm rest},{\rm rest}}\in\mathbb R_+^{(p-2)\times (p-2)}$ is the TPDM restricted to $\bs X_{{\rm rest}}$, and $\Sigma_{ik,{\rm rest}} = \Sigma_{{\rm rest},ik}^T = (\Sigma_{i,{\rm rest}}^T, \Sigma_{k,{\rm rest}}^T)^T\in\mathbb R_+^{2\times (p-2)}$ is the cross-TPDM between $\bs X_{ik}$ and $\bs X_{{\rm rest}}$. Then, based on the {projection theorem} for vector spaces with inner products, we have that 
$$\hat{\bs B} = \Sigma_{ik,{\rm rest}}\Sigma_{{\rm rest},{\rm rest}} ^{-1}.$$ 
Straightforward calculations show that the {prediction error} $\bs e = \bs X_{{ik}}\ominus\hat{\bs X}_{\text{}ik}$ has the following TPDM:
 \begin{align}\label{eq:prediction_error}
\text{TPDM}(\bs e) &= \Sigma_{ik,ik} - \Sigma_{ik,{\rm rest}}\Sigma_{{\rm rest},{\rm rest}}^{-1}\Sigma_{{\rm rest},ik}\nonumber\\
&=\begin{bmatrix}
\Sigma_{ii} - \Sigma_{i,{\rm rest}}\Sigma_{{\rm rest},{\rm rest}}^{-1}\Sigma_{{\rm rest},i} & \Sigma_{ik} - \Sigma_{i,{\rm rest}}\Sigma_{{\rm rest},{\rm rest}}^{-1}\Sigma_{{\rm rest},k} 	 \\
\Sigma_{ki} - \Sigma_{k,{\rm rest}}\Sigma_{{\rm rest},{\rm rest}}^{-1}\Sigma_{{\rm rest},i} & \Sigma_{kk} - \Sigma_{k,{\rm rest}}\Sigma_{{\rm rest},{\rm rest}}^{-1}\Sigma_{{\rm rest},k}
\end{bmatrix}.
 \end{align}

\begin{defn}\label{def:ptcc}
The \emph{partial tail correlation coefficient (PTCC)} of two random variables $X_i$ and $X_k$ is defined as the off-diagonal {TPDM coefficient of the bivariate residual vector} $\bs e$ in \eqref{eq:prediction_error}, such that transformed-linear dependence with respect to all other random variables is removed.
\end{defn}

 \begin{defn}
 Let $\bs X_{ik} = (X_i, X_k)^T$  and $\bs X_{{\rm rest}}$ be defined as above. Given $\bs X_{{\rm rest}}$, we say that ${X_i}$ and $X_k$ are \emph{partially tail-uncorrelated} if the PTCC of $X_i$ and $X_k$ (given $\bs X_{{\rm rest}}$) is equal to {zero}, i.e., if
$\Sigma_{ki}-\Sigma_{k,{\rm rest}}\Sigma_{{\rm rest},{\rm rest}}^{-1}\Sigma_{{\rm rest},i}=0$ as defined in \eqref{eq:prediction_error}.
\end{defn} 
\noindent \textbf{Remark 1}: {Thanks to the properties of TPDMs, the residuals of two partially tail-uncorrelated random variables are necessarily asymptotically tail independent.}

The following proposition links tail-uncorrelatedness to the entries of the inverse of the TPDM of $\bm X$. 
 \begin{prop}\label{prop:partial}
Given the representation of the TPDM of $\bs X'$ as a $3\times3$ block matrix as follows, 
\vspace{-10pt}
\begin{equation}\label{eq:sigma}
 	\Sigma_{\bs X'}=\text{TPDM}\{(X_i, X_k,\bm X_{{\rm rest}}^T)^T\}=
	\begin{bmatrix}
 	\Sigma_{ii} & \Sigma_{ik} & \Sigma_{i,{\rm rest}}\\
 	\Sigma_{ki} & \Sigma_{kk} & \Sigma_{k,{\rm rest}}\\
 	\Sigma_{{\rm rest},i} & \Sigma_{{\rm rest},k} & \Sigma_{{\rm rest},{\rm rest}} 
 	\end{bmatrix},
 	\end{equation}
where the dimensions of submatrices are as above, then the  following two statements are equivalent:
$$
	(1) \text{ }\Sigma_{ki} - \Sigma_{k,{\rm rest}}\Sigma_{{\rm rest},{\rm rest}}^{-1}\Sigma_{{\rm rest},i} =0,\quad\quad
	(2) \text{ }(\Sigma^{-1})_{ik}=0
$$
where $\Sigma$ is the TPDM of the original vector $\bs X$.
\end{prop}
Since $\Sigma_{\bs X'}$ is a positive definite and therefore an invertible covariance matrix, this result is a direct consequence of  the equivalence of statements $(i')$ and $(ii)$ of Proposition~$1$ of \citet{speed1986gaussian}. In the notation of \citet{speed1986gaussian}, we have the following index sets: $a=\{k,{\rm rest}\}$, $b=\{i,{\rm rest}\}$, and $ab = a\cap b = {\rm rest}$. 
The following corollary is a direct consequence of this result. 

\begin{cor}\label{prop:q}
Denote the inverse matrix of the TPDM of a random vector $\bs X$ by  $Q = \Sigma^{-1}$. Then, 
$$
Q_{ik}=0 \quad \text{if and only if}\quad \text{PTCC}_{ik}=0,
$$ 
where PTCC$_{ik}$ is the PTCC of components $X_i$ and $X_k$. {Recall that a PTCC equal to zero corresponds to partial tail-uncorrelatedness.}
 \end{cor} 
\section{Learning extremal networks for high-dimensional extremes}\label{sec:extnet}
Using the PTCC, we can now explore the partial tail correlation structure of multivariate random variables under the framework of multivariate regular variation. In this section, we define new graphical models to represent extremal dependence for extremes. Thanks to the transformed-linear framework exposed in the previous section, we can proceed as for classical graphical models by replacing the classical covariance matrix with the TPDM. 
\subsection{Graphical models for extremes}

Let $G = (V, E)$ be a graph, where $V = \{1, \ldots,p\}$ represents the node set and $E \in V\times V$ the edge set. We call $G$ an undirected graph if for two nodes $i,k\in V$, the edge $(i, k)$ is in $E$ if and only if the edge $(k,i)$ is also in $E$. We show how to estimate graphical structures for extremes for any type of undirected graph in which we have no edge $(i, k)$ if and only if the variables $X_i$ and $X_k$ are partially tail-uncorrelated given all the other variables in the graph, which we write as 
$$X_i \perp_p X_k \mid \boldsymbol{X}_{{\rm rest}}.$$

\begin{figure}[t!]
\begin{center}
    \begin{minipage}[b]{0.32\linewidth}
    \includegraphics[width=1\linewidth]{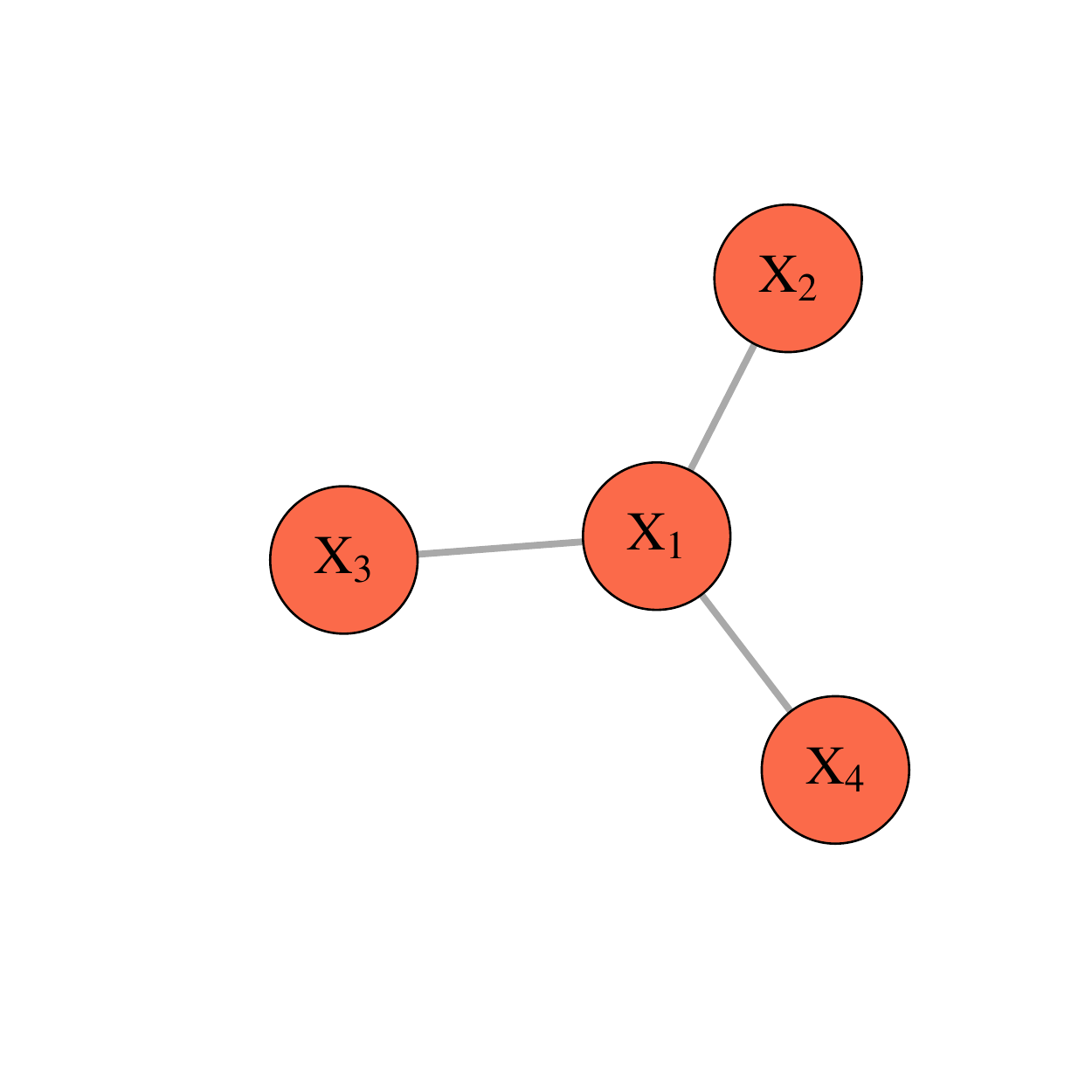} 
    \end{minipage} 
     \begin{minipage}[b]{0.32\linewidth}
    \includegraphics[width=1\linewidth]{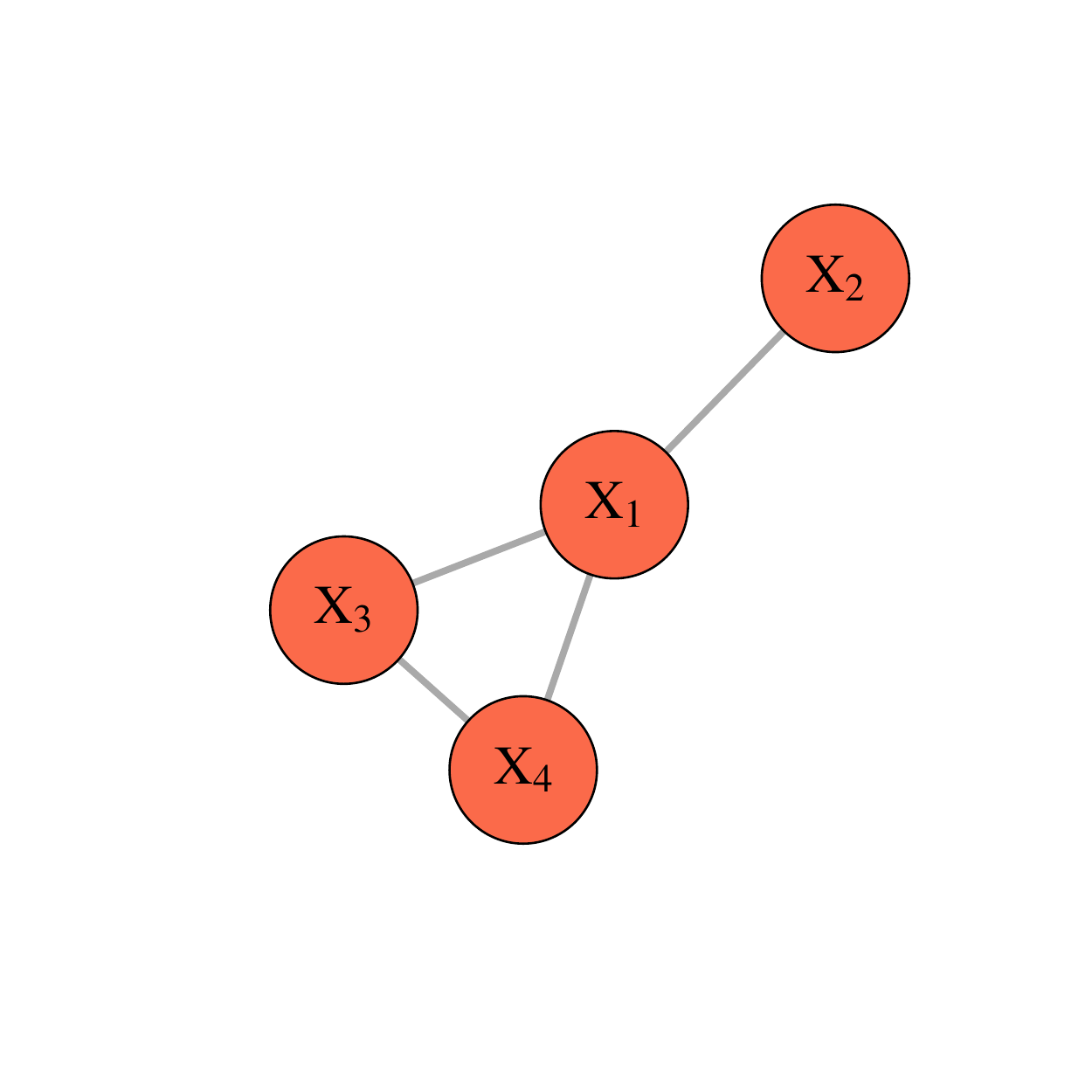} 
    \end{minipage} 
    \begin{minipage}[b]{0.32\linewidth}
    \includegraphics[width=1\linewidth]{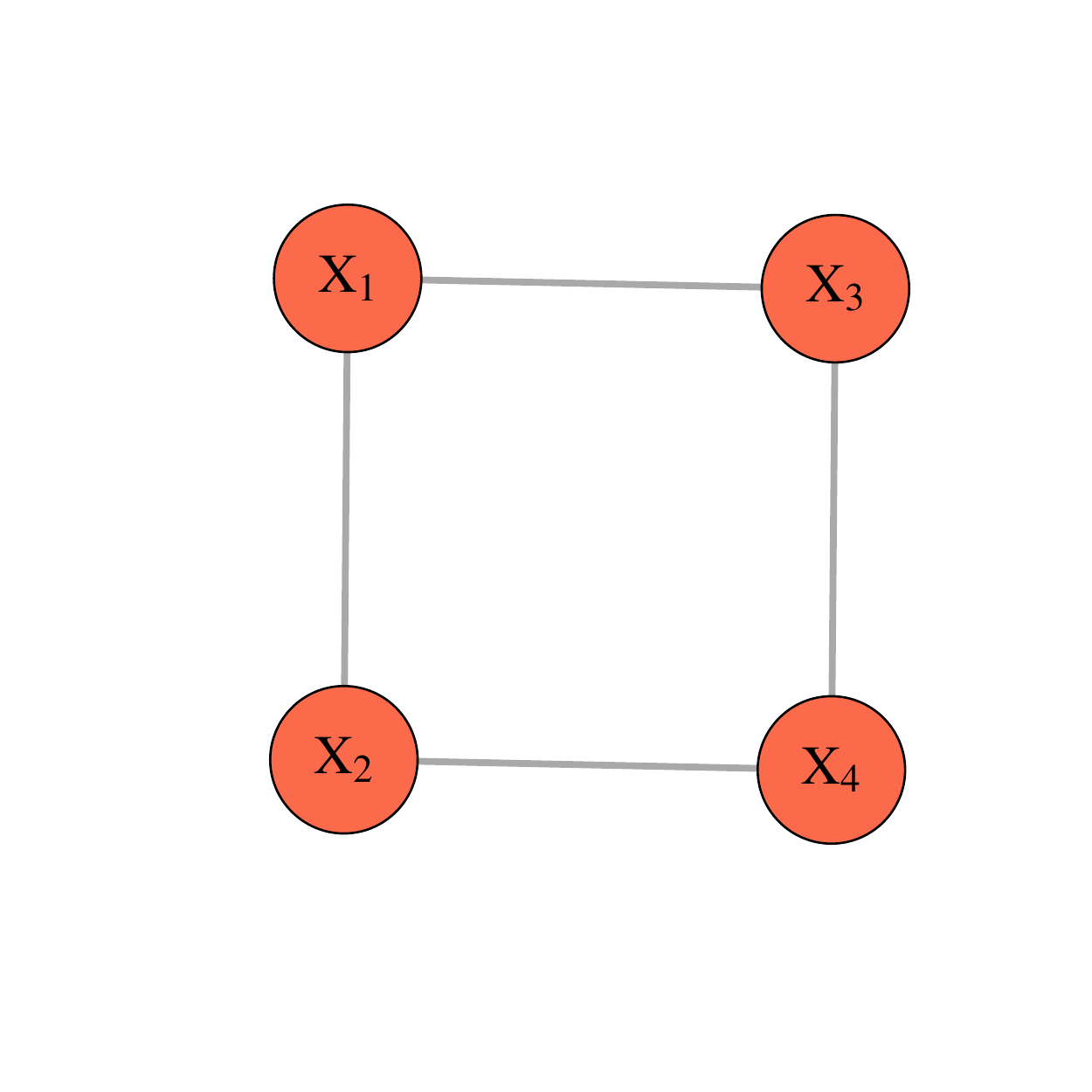} 
    \end{minipage} 
\end{center}
\caption{Examples of undirected graph structures: a tree (left), a decomposable graph (middle), and a non-decomposable graph (right).}\label{fig:graphs}
\end{figure}

Our methods work for general undirected graph structures, including trees, decomposable graphs and non-decomposable graphs, see the example illustrations in Figure~\ref{fig:graphs}. Note, however, that our general method cannot restrict the estimated graph to be of a specific type, such as a tree. 

\subsection{Sparse representation of high-dimensional extreme networks}

For high-dimensional extremes with a relatively large number of components $p$ (e.g., up to tens and hundreds of variables), a graphical representation of the extremal dependence structure is desirable for reasons of parsimony and interpretability. We now introduce two efficient inference methods to learn extremal networks from high-dimensional data via the PTCC based on two state-of-the-art graphical methods: the extremal graphical Lasso, and the structured graph learning method via Laplacian spectral constraints. These two methods both work efficiently in high-dimensional settings and return an estimate of the underlying extremal dependence with sparse structure, i.e., with the cardinality of $E$ being of the same order as the one of $V$. 

\subsubsection{Extremal graphical Lasso}\label{subsec:lasso}

Given an empirical estimator $\hat{\Sigma}$ of the TPDM and a tuning parameter $\lambda\geq 0$, the optimization carried out by the extremal graphical Lasso method is expressed as follows \citep{friedman2008sparse}:
$$
\hat{Q}_{\lambda}= \arg \max_{\Theta\succeq 0} \big[ \log\det\Theta - \text{tr}(\hat{\Sigma}\Theta) - \lambda \sum_{i\neq k}|\Theta_{ik}|\big]
$$
where $\succeq$ indicates positive-semidefiniteness and $\hat{Q}_\lambda$ is a $L_1$-regularized estimate of the precision matrix $Q=\Sigma^{-1}$. Note that thanks to the $L_1$ regularization, the estimate $\hat{Q}_{\lambda}$ will tend to be sparse (with exact zero entries) and thus contains information on the extremal graph structure. A larger $\lambda$ enforces a larger proportion of zeros in $\hat{Q}_{\lambda}$ and hence fewer edges in the graph. Choosing an appropriate value for $\lambda$ is thus critical. On the one hand, we want to enforce sparsity in the graph, where only significant connections are maintained in the network. On the other hand, $\hat{Q}_{\lambda}$ should be well-defined, with estimation being stable, and with meaningful dependence structures in the estimated model. In our river discharge application in Section~\ref{sec:appldanube} we use a voting procedure to select the best value for $\lambda$, while in our global currency application in Section~\ref{sec:applcurrency} we set the sparsity level to a pre-defined level for interpretation purposes.

\subsubsection{Structured graph learning via Laplacian spectral constraints}\label{SGL}

As an alternative to the graphical Lasso approach, we can seek to include more structural information into the graph by using the structured graph Laplacian (SGL) method of \citet{kumar2019structured}, which assumes that the signal residing on the graph changes ``smoothly'' between connected nodes. This method allows us to better balance the sparsity and connectedness of the estimated precision matrix, thanks to additional constraints on the eigenvalues of the graph Laplacian operator that encodes the graph structure. For instance, if exactly one eigenvalue is zero and all other eigenvalues are positive, then the graph is connected. Laplacian matrix estimation can be formulated as the estimation problem for a precision matrix $Q$, which is therefore linked to our framework that uses the TPDM and its inverse. For any vector of eigenvalues $\bs\lambda\in S_{\bs\lambda}$ with appropriate a priori constraints for the desired graph structure defined through the set of admissible eigenvalues $S_{\bs\lambda}$, we set {$Q = \mathcal{L} \bs w$} with $\mathcal{L}$ the linear operator that maps a non-negative set of edge weights $\bs w\in\mathbb R_+^{p(p-1)/2}$ to the matrix $Q$ with Laplacian constraints. The Laplacian matrix $Q$ can be factorized as {$Q = U{\rm Diag}(\bs \lambda)U^T$} (with an orthogonal matrix $U$) to enforce the constraints on $\bs \lambda$. Then the optimization problem can be formulated as follows:
$$
 (\hat{\bs\lambda},\hat{U}) = \arg \max_{\bs\lambda,U}\max_{\bs w} \left(
\log \text{gdet}(\text{Diag}(\bs\lambda)) - \text{tr}(\hat{\Sigma} \mathcal{L} \bs w) + \textcolor{black}{\alpha}||\mathcal{L}\bs w||_1 +\dfrac{\textcolor{black}{\beta}}{2}||\mathcal{L}\bs w - U\text{Diag}(\bs\lambda)U^T||_{F}^2
\right),
$$
$$
\text{subject to } \bs w\geq 0, \bs\lambda \in S_{\bs\lambda}, \text{and } U^TU = I,
$$
where $S_{\bs\lambda}$ denotes the set of constrained eigenvalues, $||\cdot||_F$ is the Frobenius norm, and $\text{gdet}$ is the generalized determinant defined as the product of all positive values in $\bs \lambda$. The optimization problem can be viewed as penalized likelihood if data have been generated from a Gaussian Markov random field; in more general cases such as ours, it still provides meaningful graphical structures since it can be viewed as a so-called penalized log-determinant Bregman divergence problem. Therefore, this method can be seen as an extension of the graphical Lasso that allows us to set useful spectral constraints with respect to the structure of the graph. A larger value of $\alpha$ increases the sparsity level of the graph. The hyperparameter $\beta \geq 0$ additionally controls the level of connectedness, and a larger value of $\beta$ enforces a higher level of connectedness of the estimated graph structure.
\section{Simulation study}\label{sec:simulation}
We present a simulation study with three examples where the corresponding true structures of the extremal dependence graphs are as in Figure~\ref{fig:graphs}, i.e., a tree (Case 1), a decomposable graph (Case 2), and a non-decomposable graph (Case 3). The simulated models are constructed  as follows. We simulate a dataset of $n = 10^5$ i.i.d.\ random variables $R_1, R_2, R_3, R_4\sim $ Fr\'echet$(2)$. We then construct $n$ replicates of the random vector $\bs X = (X_1, X_2, X_3, X_4)^T$  according to the following three cases, for which we also specify the true TPDM $\Sigma$ and its inverse $Q=\Sigma^{-1}$:\\
{\bf Case 1:}
\[
\begin{cases}
\notag X_1 &= R_1\\
\notag X_2 &= R_1 \oplus R_2\\
\notag X_3 &= R_1 \oplus R_3\\
\notag X_4 &= R_1 \oplus R_4
\end{cases}, 
\quad 
\Sigma= 
\begin{bmatrix}
1 &1 &1 &1\\
1 &2 &1 &1\\ 
1 &1 &2 &1\\
1 &1 &1 &2
\end{bmatrix},
\quad
Q = 
\begin{bmatrix}
4 &-1 &-1 &-1\\
-1 &1 &0 &0\\ 
-1 &0 &1 &0\\
-1 &0 &0 &1
\end{bmatrix}.
\]
{\bf Case 2:}
\[
\begin{cases}
\notag X_1 = R_1\\
\notag X_2 =R_1 \oplus R_2\\
\notag X_3 = R_1 \oplus R_3\\
\notag X_4 = R_1  \oplus 2R_3 \oplus R_4
\end{cases}, 
\quad 
\Sigma= 
\begin{bmatrix}
1 &1 &1 &1\\
1 &2 &1 &1\\ 
1 &1 &2 &3\\
1 &1 &3 &6
\end{bmatrix},
\quad
Q = 
\begin{bmatrix}
-4 &-1 &1-3 &1\\
-1 &1 &0 &0\\ 
-3 &0 &5 &-2\\
1 &0 &-2 &1
\end{bmatrix}.
\]
{\bf Case 3:}
\[
\begin{cases}
\notag X_1 = R_1 \\
\notag X_2 = R_1\oplus 3/{\sqrt{6}}R_2\\
\notag X_3 = R_1\oplus 1/\sqrt{6}R_2 \oplus 2/\sqrt{3}R_3\\
\notag X_4 = R_1  \oplus \sqrt{6}/3R_2 \oplus 1/\sqrt{3}R_3\oplus R_4
\end{cases}, 
\quad 
\Sigma= 
\begin{bmatrix}
1 &1 &1 &1\\
1 &2.5 &1.5 &2\\ 
1 &1.5 &2.5 &2\\
1 &2 &2 &3
\end{bmatrix},
\quad
Q = 
\begin{bmatrix}
2 &-0.5 &-0.5 &0\\
-0.5 &1 &0 &-0.5\\ 
-0.5 &0 &1 &-0.5\\
0 &-0.5 &-0.5 &1
\end{bmatrix}.
\]
We proceed as follows to infer the extremal graph structure in each case. First, we estimate the TPDM of $\bs X$, $\Sigma$, based on the estimator $\hat\Sigma$ specified through \eqref{eq:estimator} using the $99\%$ quantile for the threshold $r_0$ (i.e., there are $1000$ threshold exceedances to estimate the TPDM). Then, we apply the extremal graphical Lasso and SGL methods. In each setting, we test $m_1 = 300$ different values for the regularization parameter $\lambda$ when using extremal graphical Lasso and $m_2 = 400$ different settings for the combination of $\alpha$ and $\beta$ for the SGL method. The range of $\lambda$ and $\{\alpha,\beta\}$ values is chosen to span a wide range of graphical structures, from fully connected to fully sparse (no connection). In all experimental results, we have found that when the true number of edges is achieved, both methods can retrieve 100\% of the true extremal graph structure, i.e., all connections are correctly identified and the estimated graph has no wrong connections. 
 
\begin{figure}[t!]
\begin{center}
    \begin{minipage}[b]{0.24\linewidth}
    \includegraphics[width=1\linewidth]{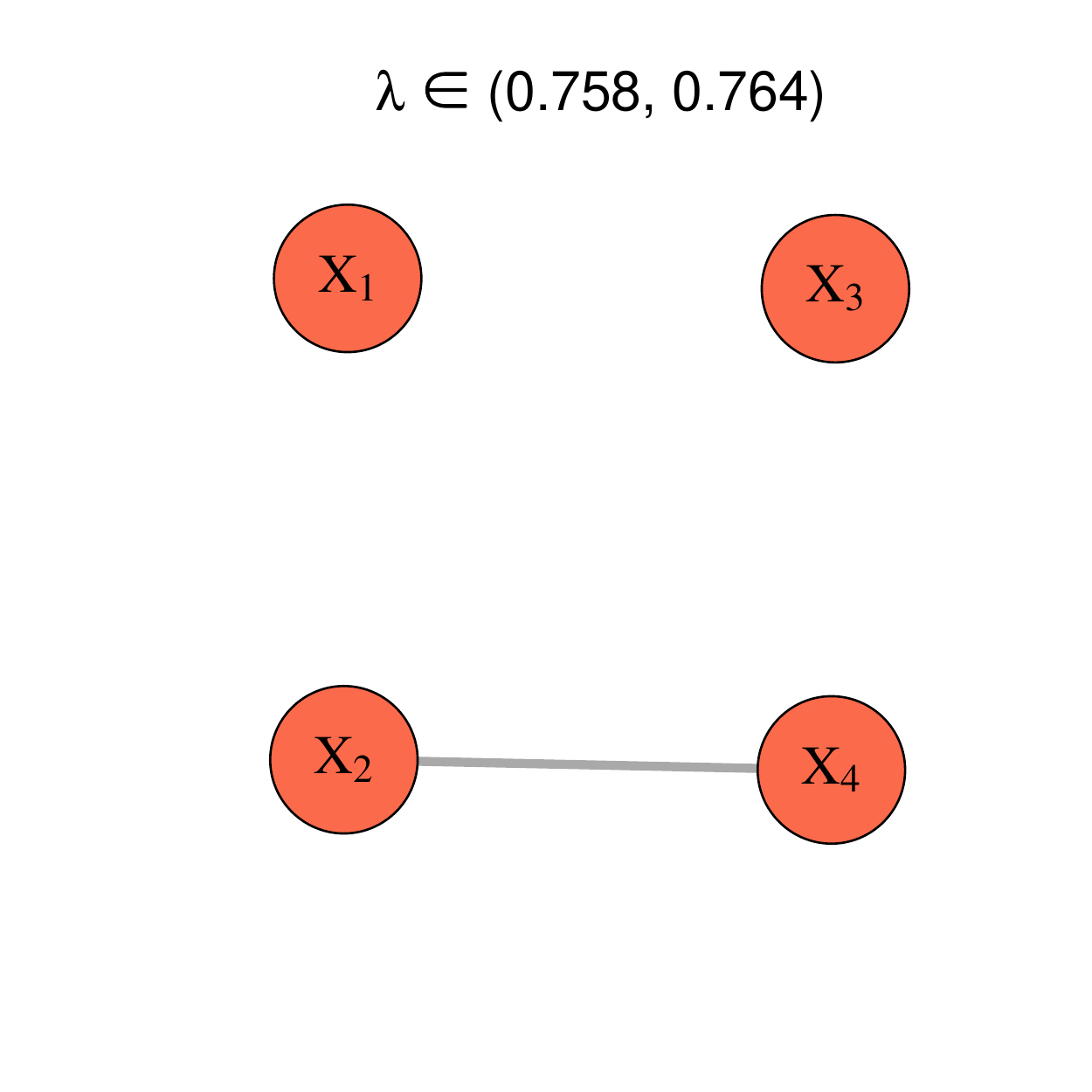} 
    \end{minipage} 
    \begin{minipage}[b]{0.24\linewidth}
    \includegraphics[width=1\linewidth]{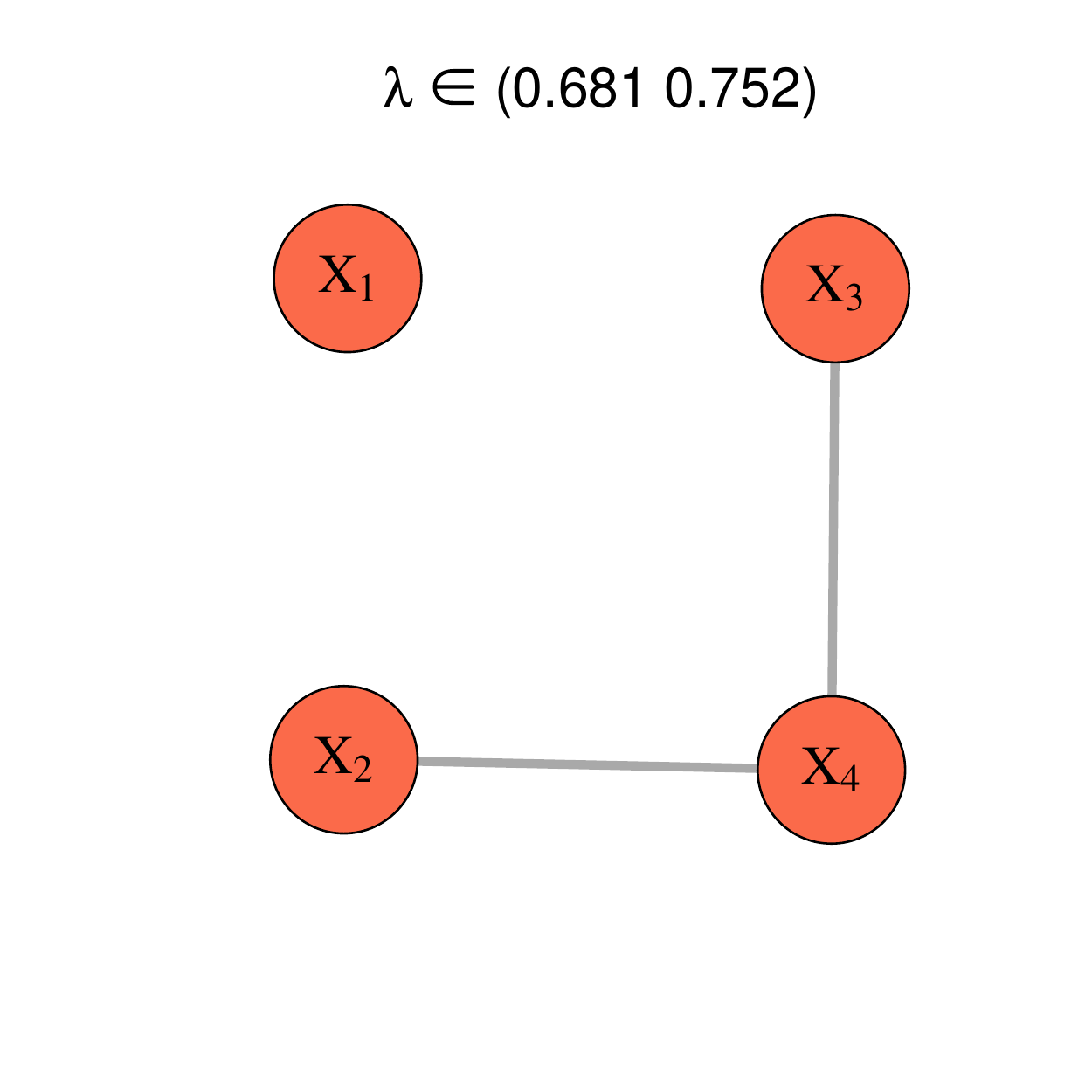} 
    \end{minipage} 
    \begin{minipage}[b]{0.24\linewidth}
    \includegraphics[width=1\linewidth]{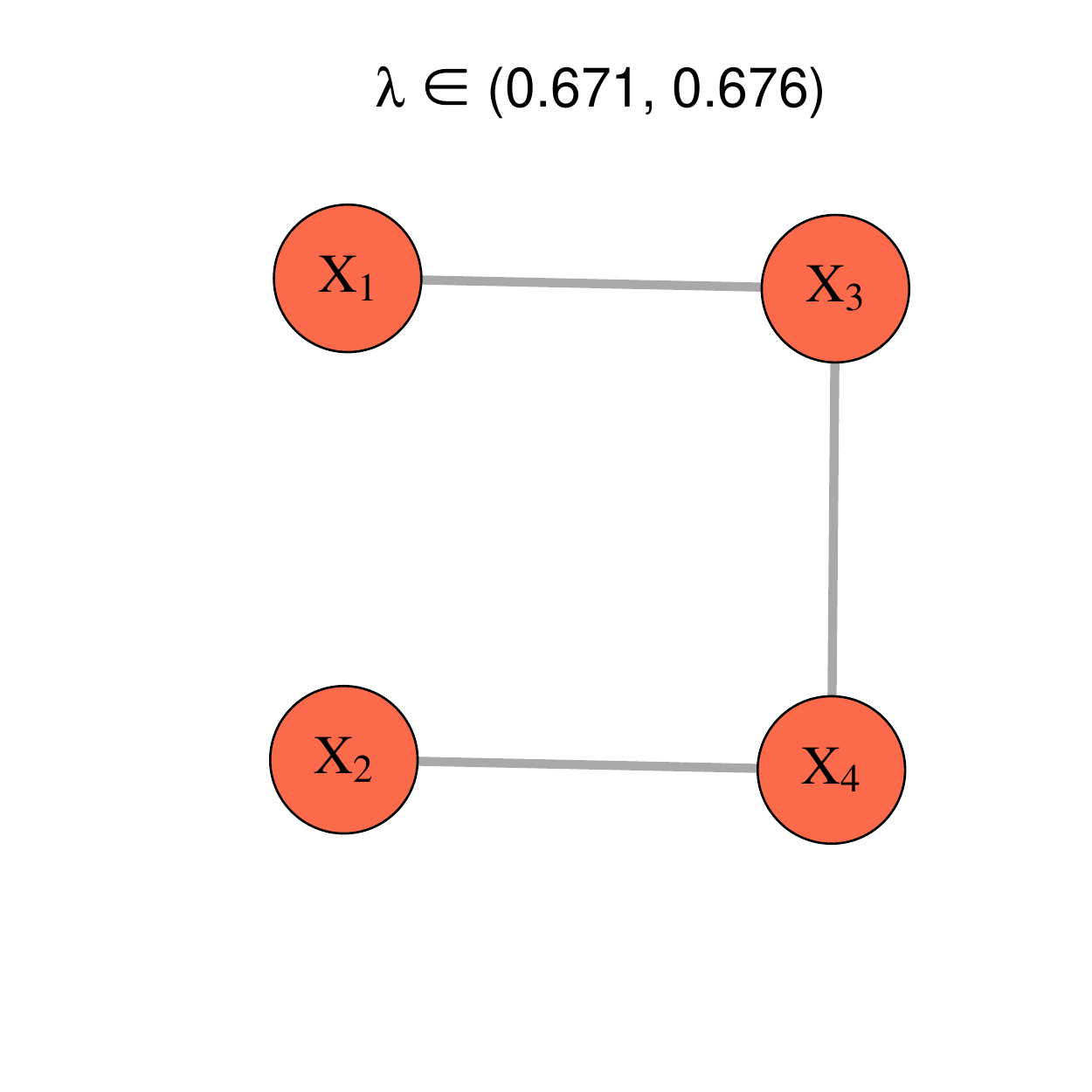} 
    \end{minipage} 
        \begin{minipage}[b]{0.24\linewidth}
    \includegraphics[width=1\linewidth]{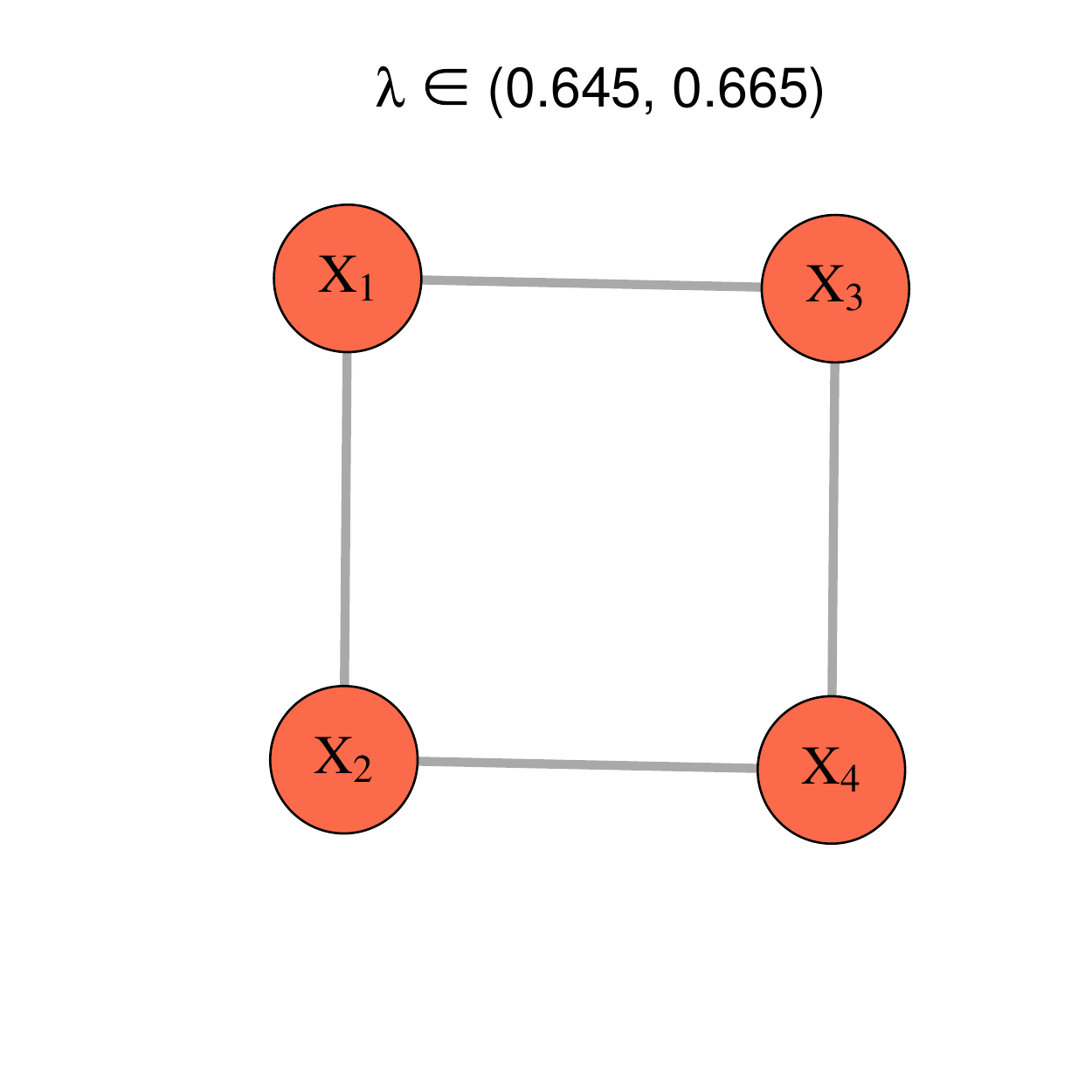} 
    \end{minipage} 
\end{center}
\caption{Estimated extremal graph structures using the extremal graphical Lasso method based on the PTCC for Case 3, as a function of the tuning parameter $\lambda$ (shown at the top of each display).}\label{fig:simu3}
\end{figure}

This is illustrated in Figure~\ref{fig:simu3}, which displays the estimated extremal graph structure for Case 3 (general non-decomposable ``square graph'') when using the extremal graphical Lasso method. The heading of each display shows the range of $\lambda$ values that leads to the estimated graph shown below. The tuning parameter $\lambda$ controls the number of edges in the graph, i.e., the sparsity level: when $\lambda$ decreases, the number of edges increases, and vice versa. Interestingly, the proposed method always retrieves true connections (i.e., it never yields wrong connections) whenever the estimated graph is as sparse, or sparser than the true graph. This simple experiment shows that our method is able to retrieve the true extremal dependence graph structure, provided the tuning parameter $\lambda$ (or $\{\alpha,\beta\}$ for the SGL method) is well specified. While our numerical experiments were performed in dimension $p=4$, we expect similar results to hold in higher dimensions provided enough data replicates are available. Our higher-dimensional data applications in Section~\ref{sec:applications} demonstrate that the estimated graph structures indeed make sense and yield interpretable results.

We also note that with our distribution-free approach, there is no universally optimal way of setting the tuning parameters. However, we can use problem-specific criteria to achieve the desired outcome (and therefore to set the penalty parameters); see the data applications in Section~\ref{sec:appldanube} and \ref{sec:applcurrency}.

 \section{Applications}\label{sec:applications}
Risk networks are useful in quantitative risk management to elucidate complex extremal dependence structures in collections of random variables. We show two examples of both environmental and financial risk analysis. First, we study river discharge data of the upper Danube basin \citep{asadi2015extremes}, which has become a benchmark dataset for learning extremal networks in the recent literature. The true underlying physical river flow network is available, which can be used as a benchmark to compare the performance of our method with other existing approaches. Second, we apply our method to  historical global currency exchange rate data from different historical periods, including different economic cycles, the COVID-19 period, and the period of the 2022 military conflict opposing Russia and Ukraine (2022.02.24--2022.09.26).
\subsection{Extremal network estimation for a river network}\label{sec:appldanube}
\begin{figure}[t!]
\begin{center}
  \begin{minipage}[b]{0.55\linewidth}
    \includegraphics[width=1\linewidth]{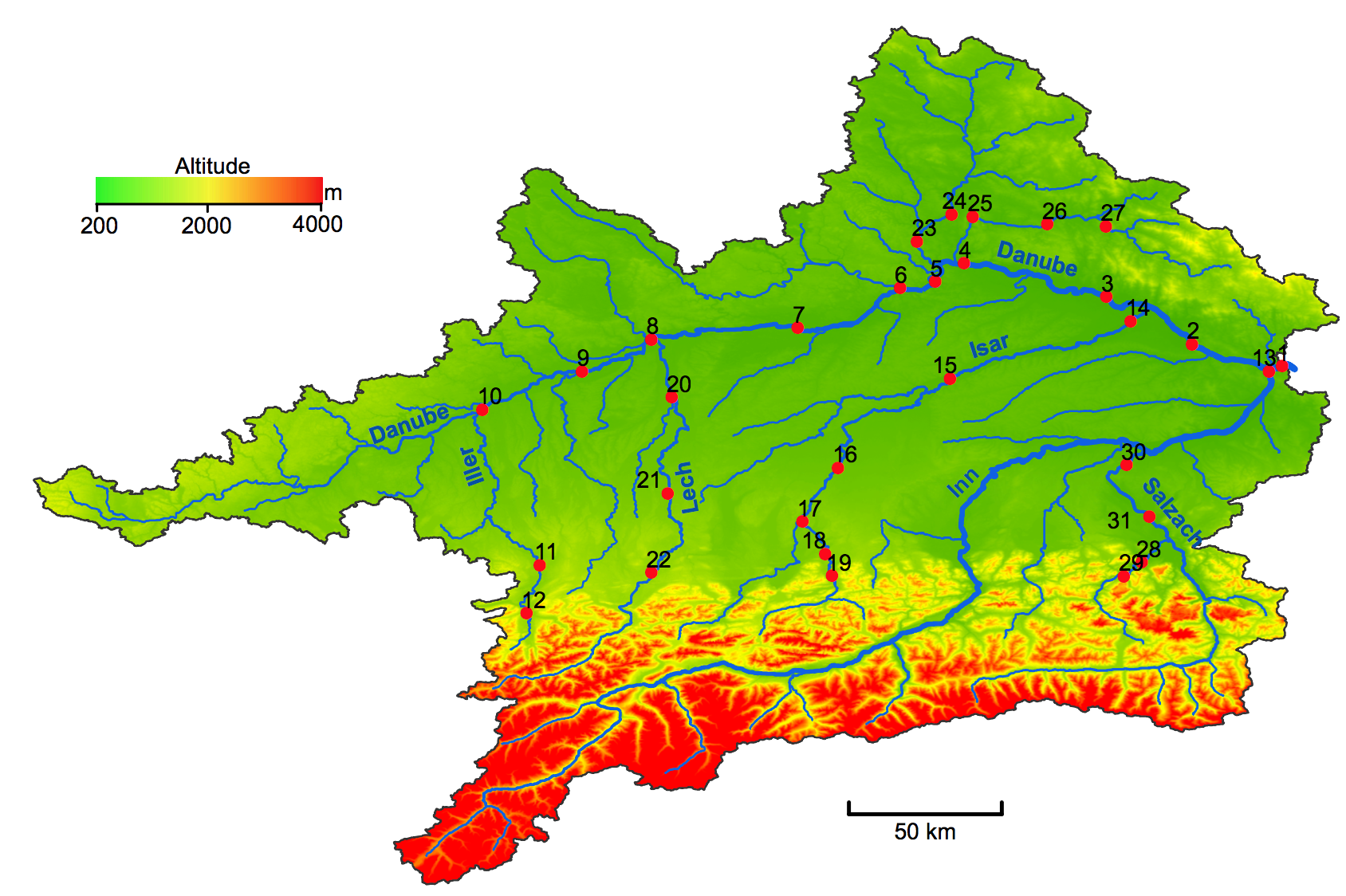} 
  \end{minipage} 
   \begin{minipage}[b]{0.4\linewidth}
    \includegraphics[width=1\linewidth]{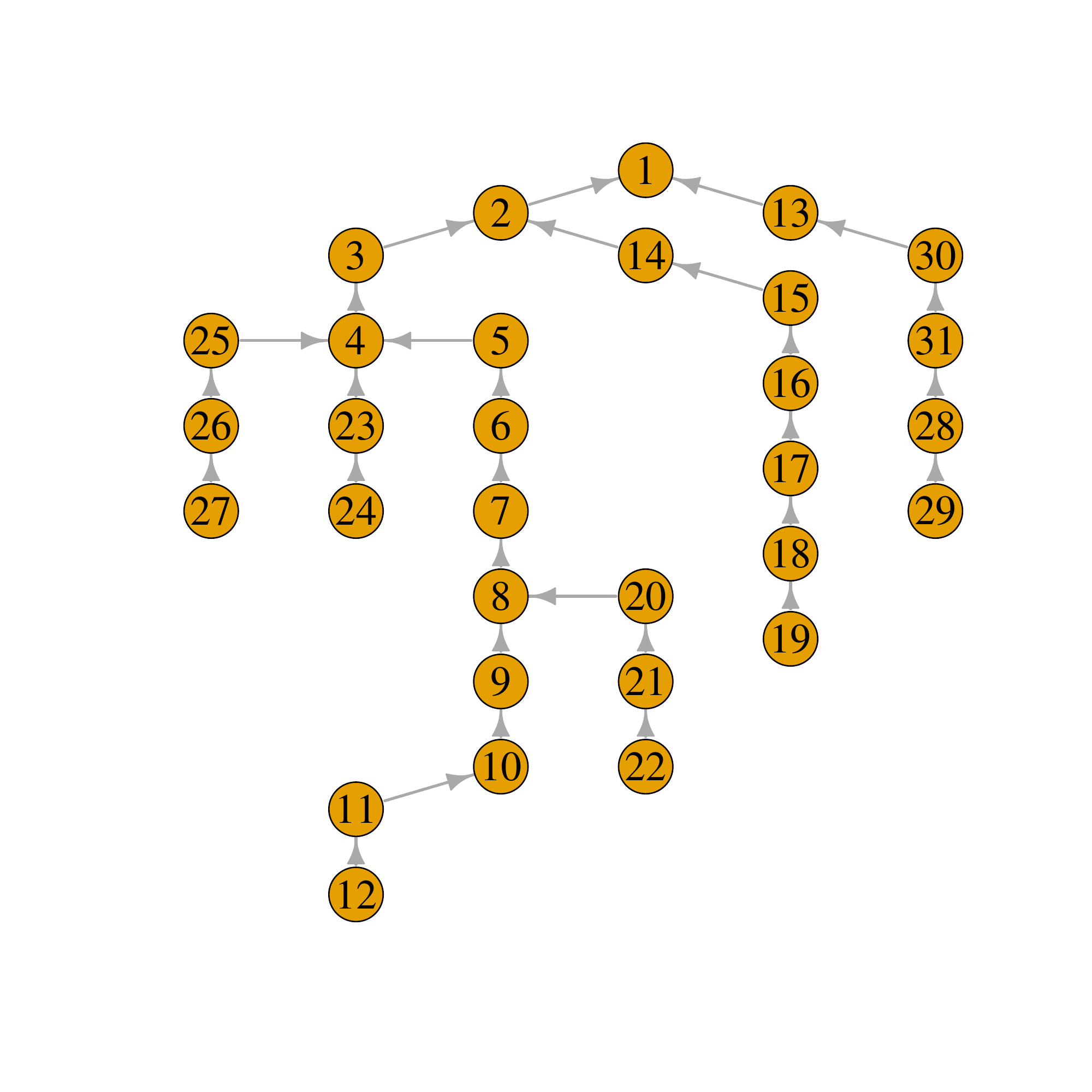} 
  \end{minipage} 
  \caption{Left: Topographic map of the upper Danube basin \citep[from][]{asadi2015extremes}, showing 31 sites of gauging stations (red circles) and the altitudes of the region. Right: The true physical river flow connections; the arrows show the flow directions.}
  \label{fig:danube}
\end{center}
\end{figure}

We apply our method to study the dependence structure of extreme discharges on the river network of the upper Danube basin (see the left panel of Figure~\ref{fig:danube} for the topographic map). This region has been regularly affected by severe flooding events in its history, which have caused losses of human lives and damage to material goods. The original daily discharge data from 1960 to 2009 were provided by the Bavarian Environmental Agency (http://www.gkd.bayern.de), and \citet{asadi2015extremes} preprocessed the data, which now include $n=428$ approximately independent events $\bs X_1,\ldots, \bs X_n \in \mathbb{R}^d$ recorded at $d=31$ gauging stations located on the river network from three summer months (June, July, and August), obtained using declustering methods. The data were later also studied by \citet{engelke2020graphical} among others, using graphical models for extremes based on a conditional independence notion adapted to multivariate Pareto distributions. The true physical river flow connections and directions are represented by a directed graph shown in the right panel of Figure~\ref{fig:danube}, where the arrows indicate the flow directions. This can serve as an accurate benchmark of the ``true'' conditional independence structure, against which we can compare the results from our proposed extremal graphical structure learning methods based on the PTCC.

\subsubsection{Graph structure learning using the extremal graphical Lasso}
\begin{figure}[t!]
\begin{center}
  \begin{minipage}[b]{0.49\linewidth}
    \includegraphics[width=1\linewidth]{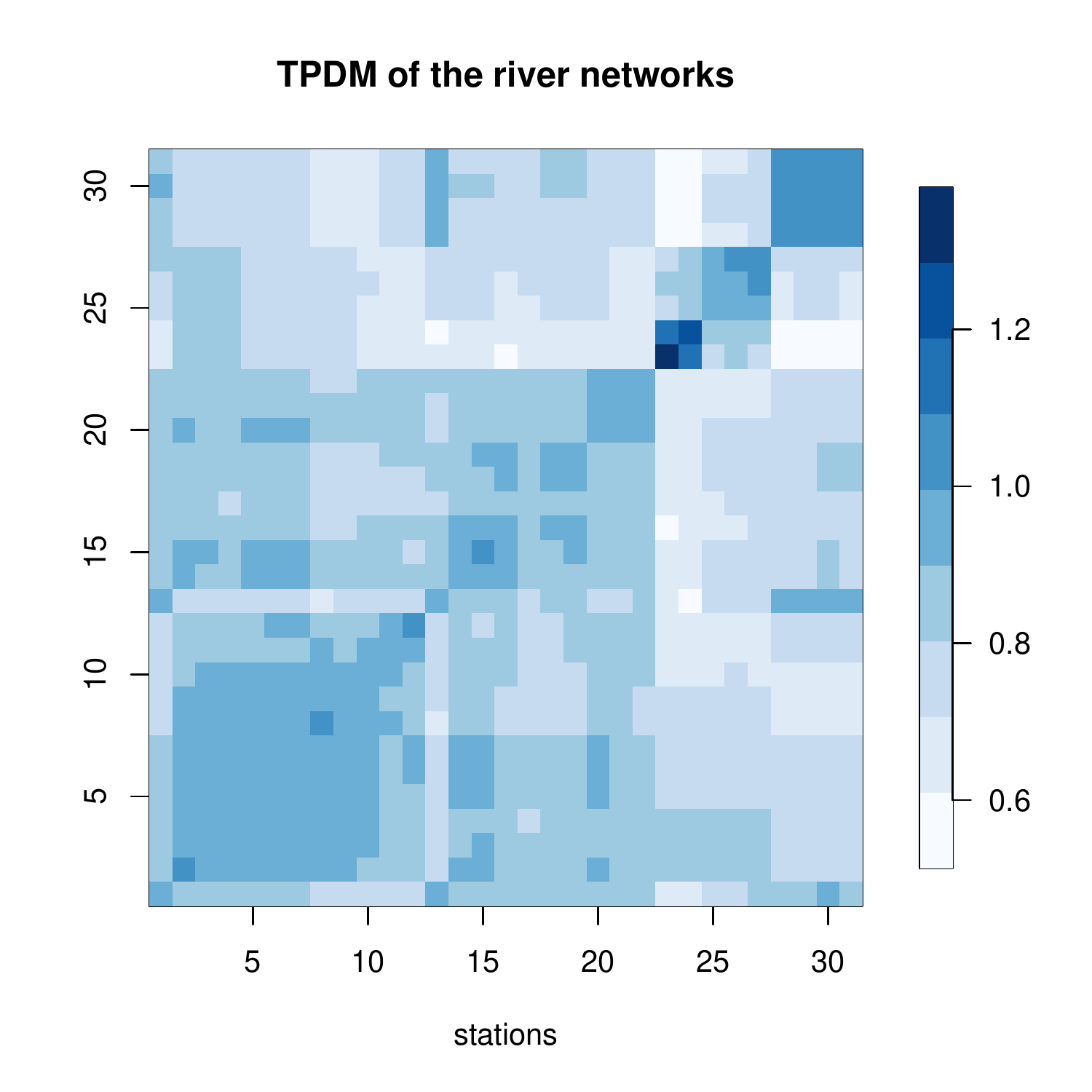} 
  \end{minipage} 
   \begin{minipage}[b]{0.49\linewidth}
    \includegraphics[width=1\linewidth]{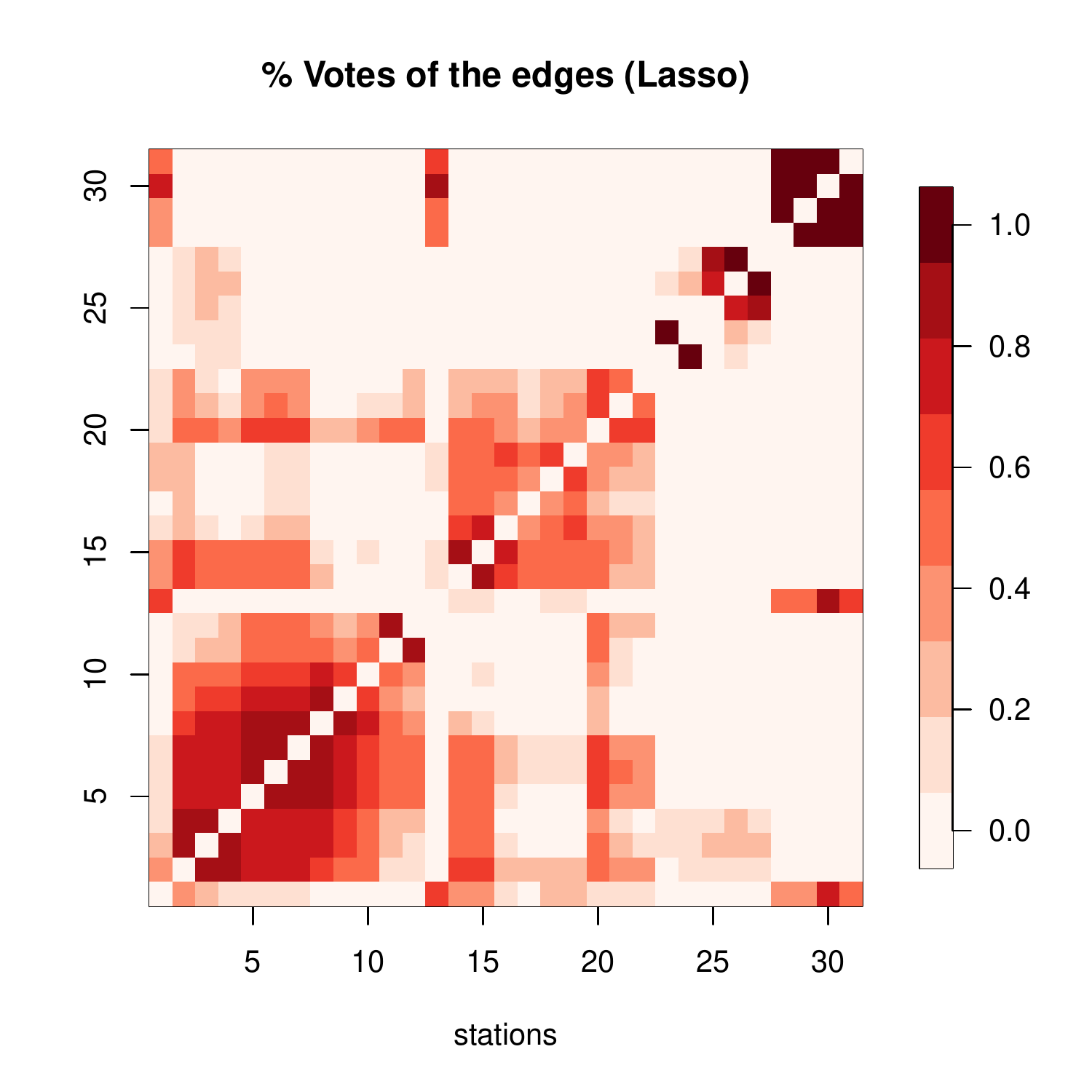} 
  \end{minipage} 
  \caption{Left: Estimated TPDM of the river discharge data from the upper Danube basin. Right: Votes ($\%$) of the edges selected based on the extremal graphical Lasso method. Darker red cells indicate that the corresponding edge has been selected more often by the graphical Lasso.}
  \label{fig:tpdm}
\end{center}  
\end{figure}

To learn the extremal dependence structure of the river network, we first perform a nonparametric empirical transformation of the data to satisfy Fr\'echet($\alpha = 2$) margins, for each station separately. Next, we estimate the TPDM, $\Sigma$, using the proposed estimator, $\hat\Sigma$, defined through \eqref{eq:estimator}. In particular, we choose $m=d = 31$ because the margins are preprocessed to have a common unit Fr\'echet scale and $r_0 = 11.4$, which corresponds to the empirical 90\% quantile. Therefore, $n_{\text{exc}} = 43$ is the number of extreme observations (i.e., threshold exceedances) which are used to estimate $\Sigma$. The left panel of Figure~\ref{fig:tpdm} displays the estimated TPDM of the river discharge data from the upper Danube basin, while the right panel displays the votes (in percentage) of the edges selected based on the extremal graphical Lasso method, obtained from multiple fits with a range of $\lambda$ values producing different dependence structures, from fully connected to fully sparse graphs.

\subsubsection{Graph structure learning using the SGL method}
\begin{figure}[t!]
\begin{center}
  \begin{minipage}[b]{0.48\linewidth}
    \includegraphics[width=1\linewidth]{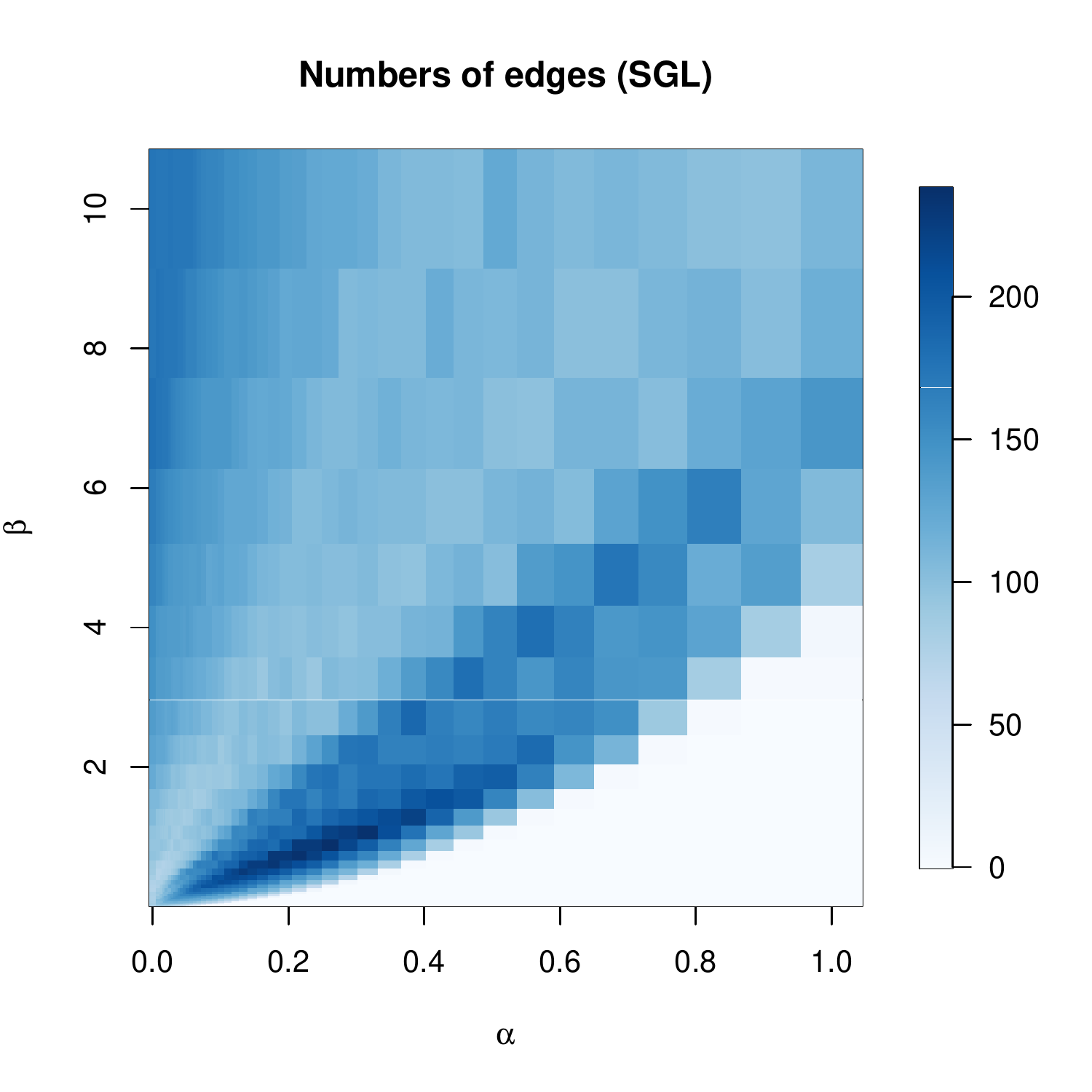} 
  \end{minipage} 
     \begin{minipage}[b]{0.48\linewidth}
    \includegraphics[width=1\linewidth]{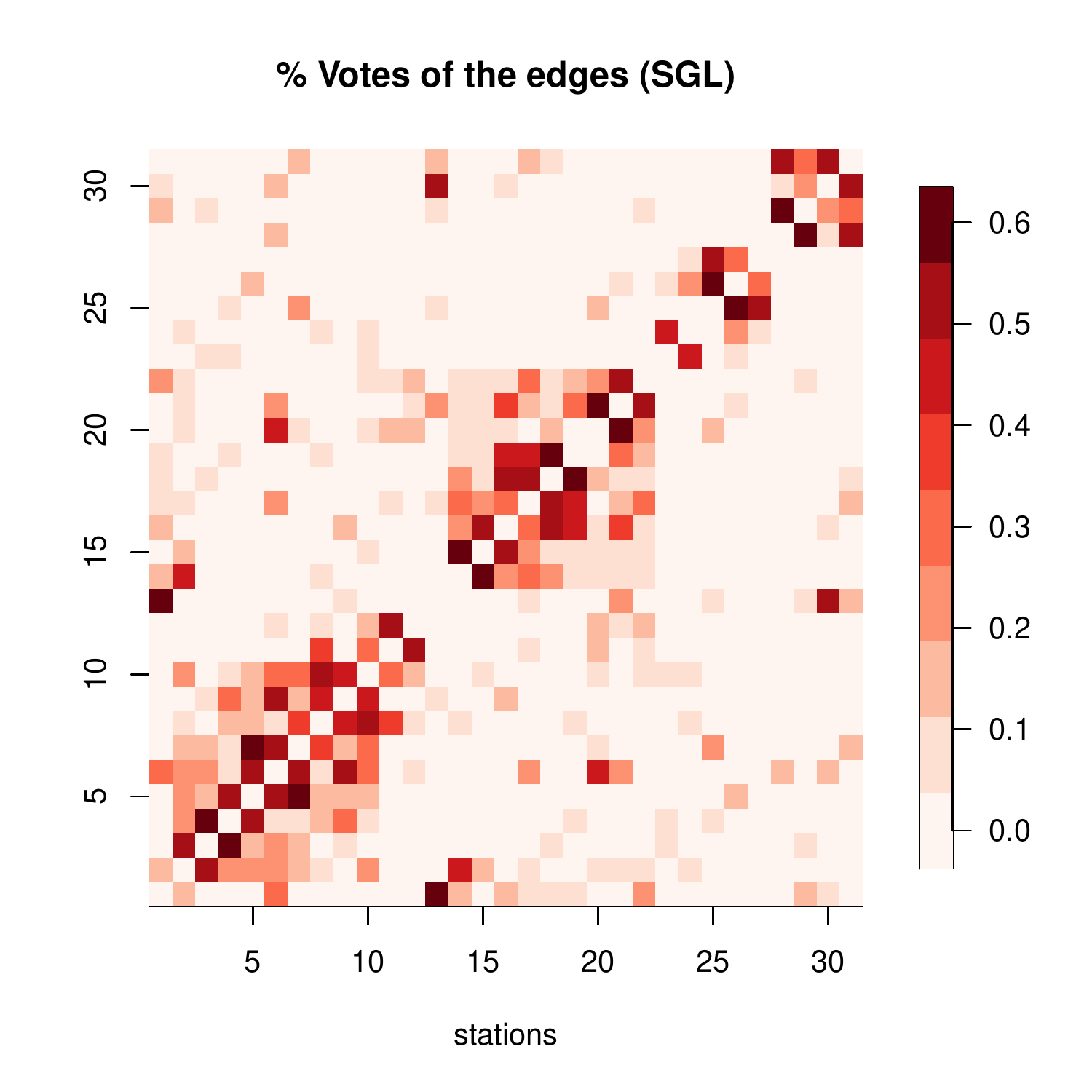} 
  \end{minipage} 
  \caption{Left: Number of edges selected by the SGL method as a function of the tuning parameters $\alpha$ and $\beta$. Lighter blue cells correspond to parameter combinations producing sparser graphs. Right: Votes ($\%$) of the edges selected based on the SGL method. Darker red cells indicate that the corresponding edge has been selected more often by the SGL method.}
  \label{fig:SGL}
\end{center}
\end{figure}

To enhance connectedness, we further explore the SGL method, which learns sparse graph structures under additional spectral constraints. In particular, as described in Section~\ref{SGL}, we can control both the sparsity level and the graph connectedness by modulating the two tuning parameters $\alpha$ and $\beta$, respectively. As shown in the left panel of Figure~\ref{fig:SGL}, the overall graph sparsity varies for different combinations of $\alpha$ and $\beta$. The right panel of Figure~\ref{fig:SGL} displays the votes (in percentage) of the edges selected based on the SGL method, obtained by fitting a large number of models for each of the $\{\alpha,\beta\}$ combinations shown in the left panel.  

\subsubsection{Results: estimated extremal river discharge networks}

For both the extremal graphical Lasso and the SGL method, it is important to carefully select the tuning parameters, $\lambda$ and $\{\alpha,\beta\}$, respectively. These tuning parameters impact both the sparsity level and the connectedness of the resulting graph structure. Ideally, we would like to obtain a sparse graph while keeping it connected. However, there is a tradeoff between these two requirements. Our approach is to control the overall sparsity level while imposing a soft connectedness condition: we start from the fully sparse graph (no edges) and sequentially add edges between nodes according to the ranking of the votes shown in Figures~\ref{fig:tpdm} and \ref{fig:SGL}, until no node is left alone (i.e., each node has at least one connection with another node). The estimated graph structure thus prioritizes edges that are most often selected and is obtained by blending the results from several model fits, which also makes it less sensitive to specific values of the tuning parameters. 

\begin{figure}[t!]
\begin{center}
  \begin{minipage}[b]{0.49\linewidth}
    \includegraphics[width=1\linewidth]{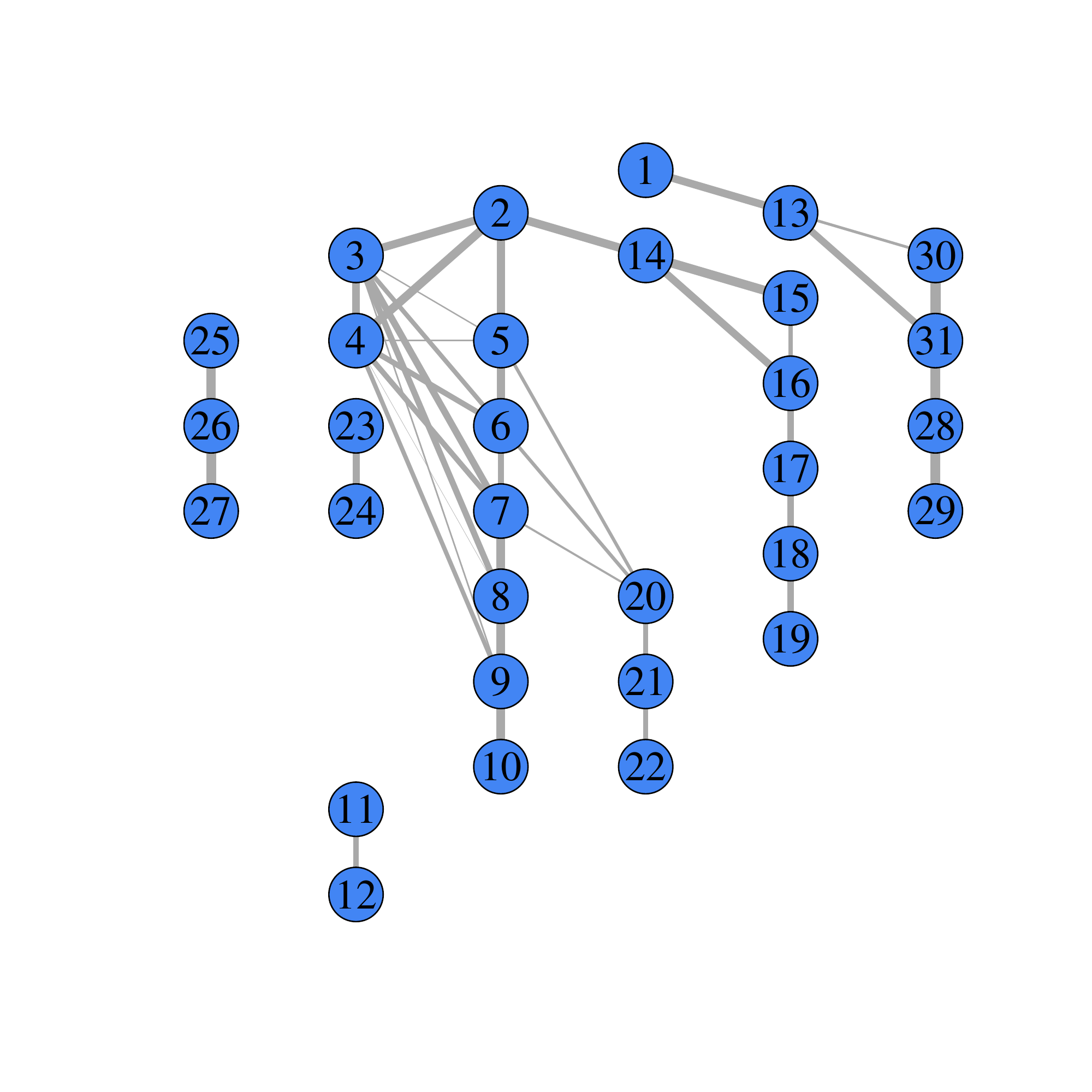} 
  \end{minipage} 
   \begin{minipage}[b]{0.49\linewidth}
    \includegraphics[width=1\linewidth]{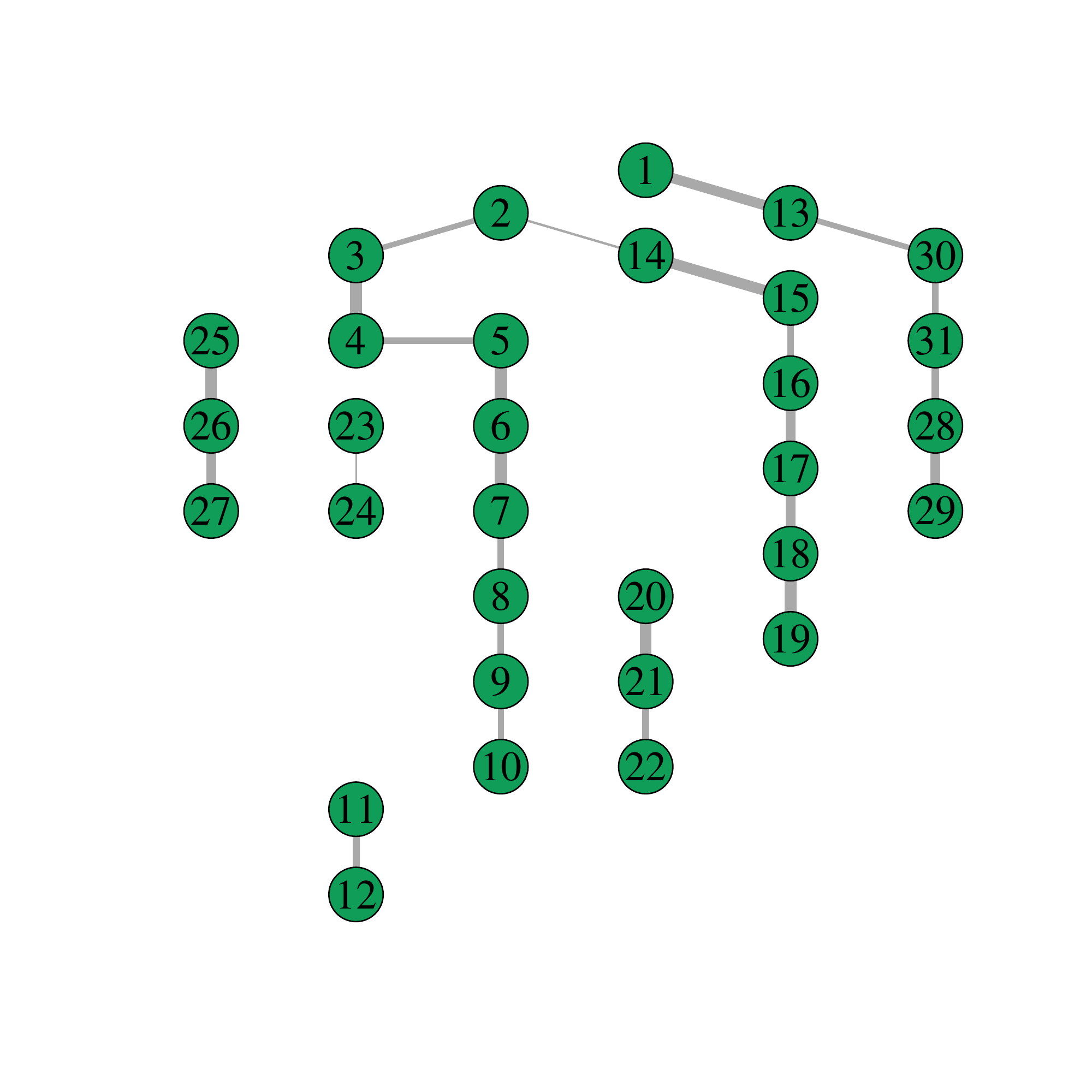} 
  \end{minipage} 
  \caption{Extremal river discharge network estimated using the extremal graphical Lasso (left) and the SGL method (right). The edge thickness is proportional to the votes shown in Figures~\ref{fig:tpdm} and \ref{fig:SGL}.}
  \label{fig:river_result}
\end{center}
\end{figure}

Figure~\ref{fig:river_result} displays the estimated extremal river discharge network using our approach for both the extremal graphical Lasso and the SGL method, respectively. The edge thickness is proportional to the votes shown in Figures~\ref{fig:tpdm} and \ref{fig:SGL}.
Recall that when two nodes are not connected, it means that they are partially tail-uncorrelated in our framework. The estimation based on the extremal graphical Lasso method has a few more edges than the true physical river flow network. The extra links could be interpreted as being due to extremal dependence induced by regional weather events, though this is not very clear. By contrast, the estimation result based on the SGL method matches most of the true river flow connections, while the votes (shown in terms of the edge thickness) represent the strength of the extremal dependence connections. Interestingly, this dependence strength seems well-aligned with the actual physical strength of the river flow. Compared with existing methods \citep{engelke2020graphical,  kluppelberg2021estimating}, our estimated extremal network based on the SGL method looks more realistic (thus more easily interpretable), as it is closer to the true flow structure of the river network, though the recent results from \citet{Engelke.etal:2022} are quite similar to ours (with a few extra connections).

\subsection{Extremal network estimation for global currency exchange rate network}\label{sec:applcurrency}

We now apply our method to explore historical global currency exchange rate data for 20 currencies from different historical periods, including two different global economic cycles (2009--2014 and 2015--2019, where the segmentation is determined from the world GDP cycles illustrated in Figure~\ref{fig:gdp} from Appendix~\ref{appd:currency}), COVID-19 (2020.01.01--2022.02.23), and the period from the beginning of the 2022 military conflict between Russia and Ukraine until a most recent date when we downloaded the data (2022.02.24--2022.09.26). Historical data were downloaded from \href{https://finance.yahoo.com/?guccounter=1&guce_referrer=aHR0cHM6Ly93d3cuZ29vZ2xlLmNvbS8&guce_referrer_sig=AQAAAArrJ51ii0LwZtFYwOr7XyQJJJefzGD3DqIN_OCi-CjH2j981I5L7tgeoZv4eTXCY0OSAxXw09alMxpKV7ygEdoDki02tMTic03tGlae7MXXVHzLszZQgvx5L9EFfqzlYo-S6UGPY7IHtZbO6fi1ZFbs_5hcGAkjzmw24f-EE8KR}{Yahoo Finance}. We chose the currencies from all G20 countries, and also added the Ukrainian and Kazakhstani currencies. The list of selected currencies and their corresponding symbols can be found in Table~\ref{table:currency} in Appendix~\ref{appd:currency}. Since the unit of the currencies is the US Dollar (USD), USD is not considered in our list of currencies under study. 

First, we preprocess the historical daily closing prices of the currencies. An ARMA(1,1)-GARCH(1,1) time series model is fitted to the negative log return time series of each currency, and then standardized residuals are extracted and transformed marginally to Frech\'et margins with shape parameter $\alpha=2$. The extremal dependence graph structure governing negative log returns therefore represents partial associations among extreme losses, shedding light on the integration and/or vulnerability of major economies in periods of stress. To estimate this risk network, we follow the same procedure as before, estimating first the TPDM using the estimator in \eqref{eq:estimator} with $r_0$ as the empirical $90\%$ quantile, and then applying the SGL method.

\begin{figure}[t!]
\begin{center}
  \begin{minipage}[b]{0.49\linewidth}
    \includegraphics[width=1\linewidth]{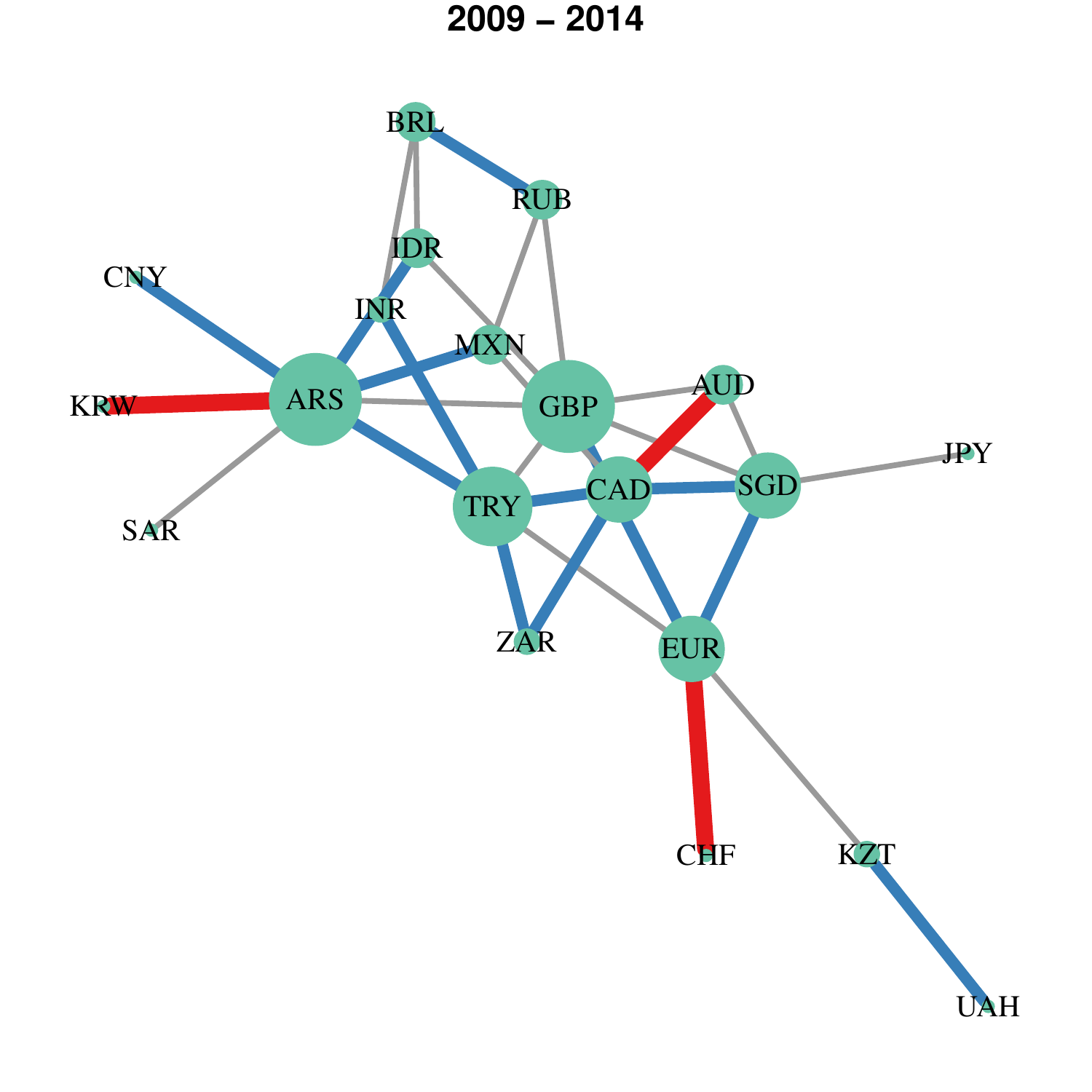} 
  \end{minipage} 
   \begin{minipage}[b]{0.49\linewidth}
       \includegraphics[width=1\linewidth]{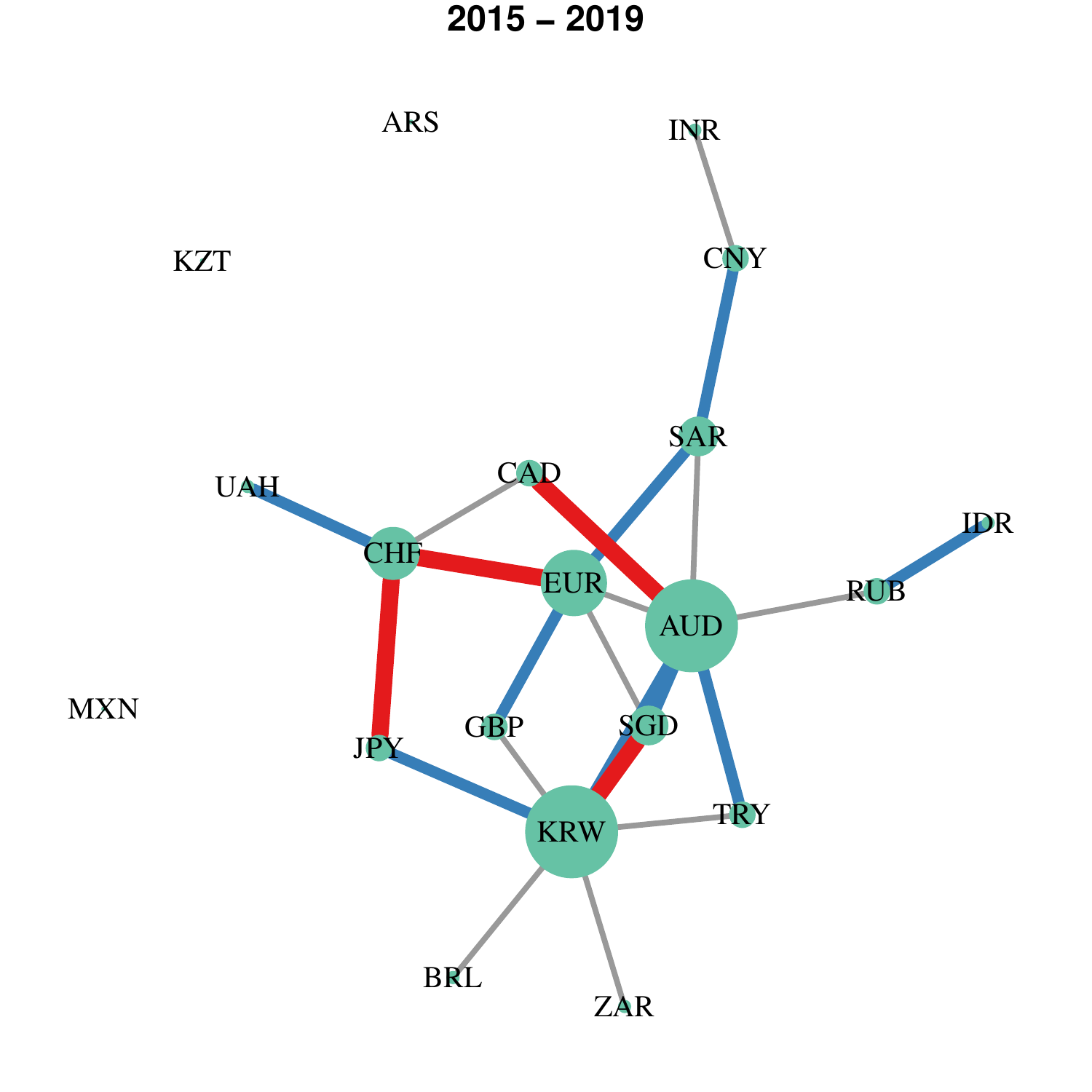} 

  \end{minipage} 
  \caption{Extremal currency exchange rate risk network estimated for 2009--2014 (left) and 2015--2019 (right). Different edge types (respectively thickest, thick, and thin) with different colors (respectively red, blue, and grey) indicate a frequency of $>90\%$, $70\%$--$90\%$, and $50\%$--$70\%$, respectively, to be selected among the $300$ bootstrap fitted models, while absent edges indicate that this frequency is less than $50\%$.}
  \label{fig:eco}
\end{center}
\end{figure}

\begin{figure}[t!]
\begin{center}
  \begin{minipage}[b]{0.49\linewidth}
   \includegraphics[width=1\linewidth]{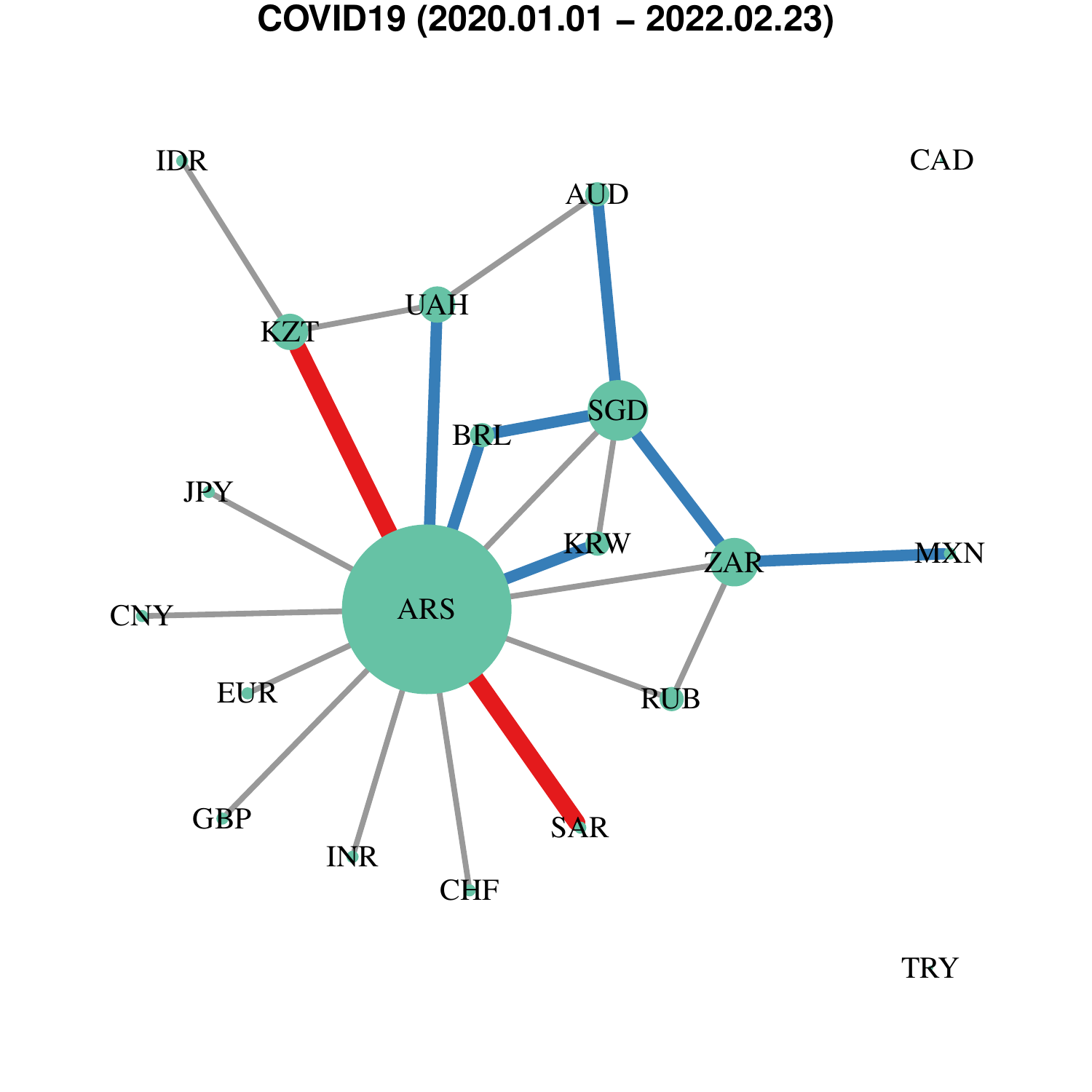}
  \end{minipage} 
   \begin{minipage}[b]{0.49\linewidth}
    \includegraphics[width=1\linewidth]{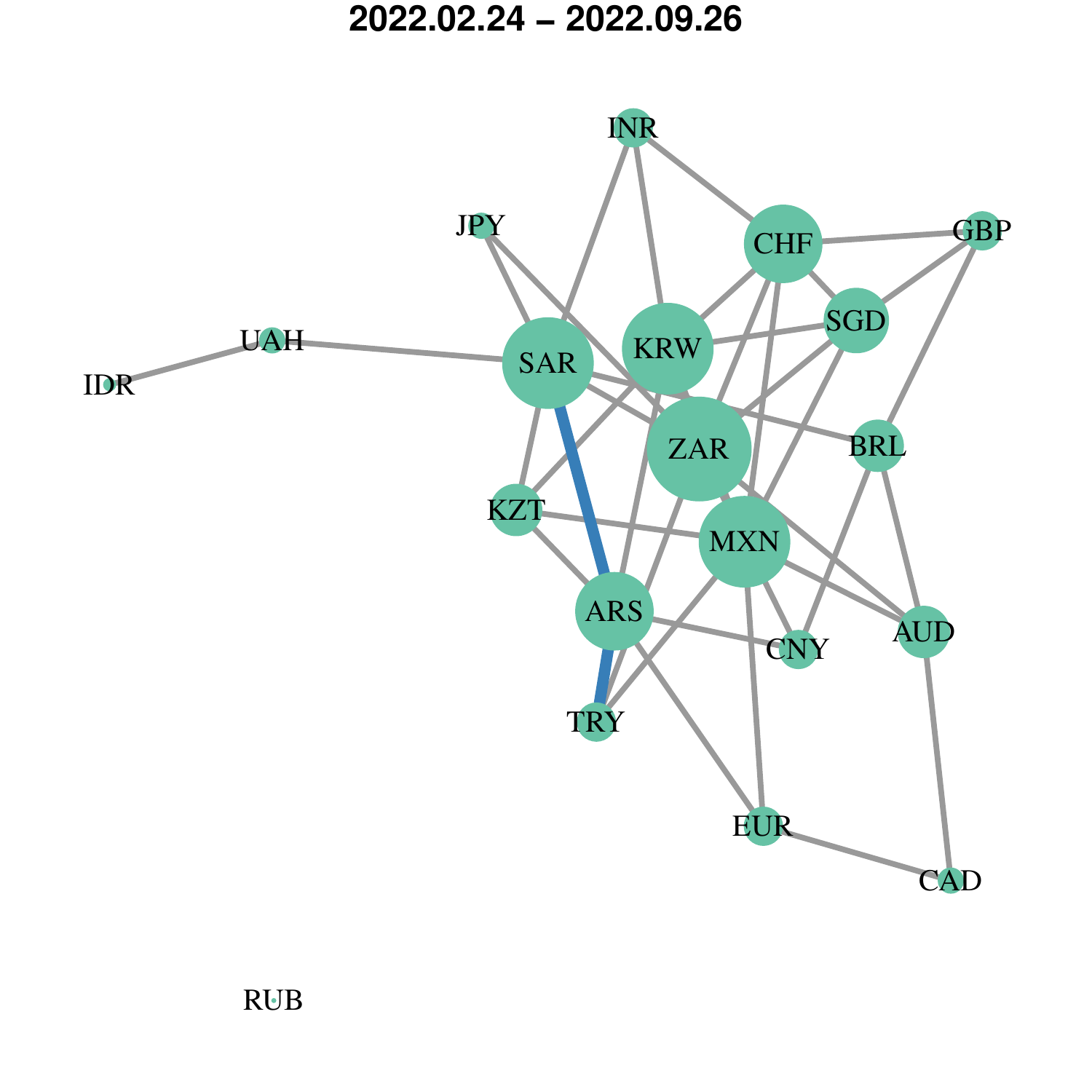} 
  \end{minipage} 
  \caption{Extremal currency exchange rate risk network estimated for the COVID-19 period (left) and the period since the 2022 military conflict between Russia and Ukraine (right). Different edge types (respectively thickest, thick, and thin) with different colors (respectively red, blue, and grey) indicate a frequency of $>90\%$, $70\%$--$90\%$, and $50\%$--$70\%$, respectively, to be selected among the $300$ bootstrap fitted models, while absent edges indicate that this frequency is less than $50\%$.}
  \label{fig:events}
\end{center}
\end{figure}

In this analysis, we use only the SGL method, because it includes the extremal graphical Lasso method as a special case when $\beta = 0$, and it has shown better performance in the Danube river application. Furthermore, for comparability among different historical periods, we here fix the sparsity level to 80\% (i.e., with only about 38 edges), rather than using the method based on votes. This approach yields a unique combination of $\{\alpha,\beta\}$ tuning parameters, which then yields the final estimated graph structure. To assess the estimation uncertainty of the graph structure, we have further conducted 300 bootstrap simulations, whereby standardized residuals are resampled with replacement and the TPDM is then re-estimated, as well as the graph structure fixing the same sparsity level (i.e., potentially with different selected $\{\alpha,\beta\}$ values for each bootstrap simulation). The bootstrap results are shown in Figures~\ref{fig:eco} and \ref{fig:events} for the first two, and last two periods, respectively, where different edge types represent the ``significance'' of the displayed connections: the thickest edges (in red) indicate a frequency of at least $90\%$ to be included among the 300 bootstrap fitted models; thick edges (in blue) indicate a frequency between $70\%$ and $90\%$ to be selected; thin edges (in grey) indicate a frequency between $50\%$ and $70\%$ to be selected; absent edges have been selected less than $50\%$ of the time. Moreover, the size of nodes in the displayed graphs is proportional to their degree, i.e., to their number of connections. The bigger a node, the more connected it is, giving an idea of the centrality of a currency in the risk network.

We now provide some interpretation of the estimated risk networks, assuming that more strongly connected nodes tend to be more vulnerable/exposed to network risks, or are important currencies that strongly determine the behavior of other currencies in the monetary system.  We can back up some of our findings using historical events and economic development regimes for different countries. From the left panel in Figure~\ref{fig:eco}, representing the economic cycle 2009--2014, the strongest-connected currencies within the estimated risk network are GBP (United Kingdom) and ARS (Argentina), whereas KZT (Kazakstan) and UAH (Ukraine) are less connected from the major clusters. By contrast, in the right panel of Figure~\ref{fig:eco}, representing the economic cycle 2015--2019, the strongest-connected currencies from the estimated risk network are KRW (Republic of Korea), AUD (Australia) and EUR (EU), whereas ARS, KZT, and MXN (Mexico) are relatively isolated. ARS and GBP are two examples of currencies that were much weaker connected during the second period. Historical economic data show that the Argentinian economy went through a major crisis around 2000, establishing strong economic links to other countries and international financial markets in the following years. This seems to clearly transpire from the estimated graph for the 2009--2014 period following the global financial crises of 2008, but the monetary interconnections of ARS to other currencies have drastically been reduced in the later 2015--2019 period. As to GBP, the exit of Great Britain from the European Union (Brexit), and the political and economic decisions accompanying it, could be the origin of this change.  
The left panel of Figure~\ref{fig:events} shows that during the COVID-19 pandemic, the ARS currency is again the strongest-connected one, by far, in the risk network, whereas TRY (Turkey) and CAD (Canada) are isolated and most of the other currencies are also less interconnected. As to ARS, Argentina was struck by  COVID-19 during an already fragile economic situation, and its aforementioned strong international monetary ties might have led to strong extremal connectedness to other countries. Finally, the right panel of Figure~\ref{fig:events} shows the risk network for currency exchange rates from the beginning of the 2022 military conflict between Russia and Ukraine until 2022.09.26. Since the length of the time period is shorter, there is higher estimation uncertainty, and no edges have been selected more than $90\%$ of the time among the $300$ bootstrap model fits. Nevertheless, we can still make the interesting observation that RUB (Russia) is the only isolated currency from the rest of the network, which might be due to the antagonism and strong economic sanctions imposed by western countries, as well as the dramatic changes in monetary policies.

\section{Conclusions}\label{sec:conclusion}
We have proposed the partial tail correlation as a novel notion of extremal dependence measure that removes the effect of confounding variables through transformed-linear operations. Unlike other approaches from the recent literature, our new partial tail correlation coefficient (PTCC) assumes multivariate regular variation but does not rely on any further strict parametric assumptions. Furthermore, the PTCC has appealing theoretical properties and it can be used to define a new class of extremal graphical models, where the absence of edges indicates partial tail-uncorrelatedness between variables (i.e., when the PTCC of the corresponding edges equals zero). We have shown that the zero PTCC values between variable pairs can be retrieved by identifying the zero entries in the inverse tail pairwise dependence matrix (TPDM). This convenient property, which is akin to classical Gaussian graphical models, allows us to efficiently learn high-dimensional extremal networks defined in terms of the PTCC by exploiting state-of-the-art methods from graph theory, such as the graphical Lasso and structured graph learning via Laplacian spectral constraints. 


Our graph-inference approach is flexible, can be applied to general undirected graphs, and easily scales to high dimensions. We demonstrate the effectiveness of our method as an exploratory tool for interpretable extremal network estimation. In our first application to river discharge data from the upper Danube basin, we show that the proposed method outperforms other existing methods by realistically capturing most physical flow connections, together with the strength of the connections, while largely avoiding spurious connections. In our second application to historical currency exchange rate data, we obtain interesting findings based on the estimated risk network for four recent periods, which can be backed up by real historical evidence. While our interpretations remain fairly basic, it would be interesting to get further insights from economists.

We have identified some theoretical and methodological challenges for future research. First, we have not obtained theoretical guarantees that our graph learning method can consistently estimate the unknown graph structure, and it would be interesting to see if this could be proven under general assumptions. However, our simulations and applications have provided convincing evidence that the method works well to extract useful structural information from extreme observations. Moreover, it is also worth noting that in some cases, there is no interest in recovering the exact true graph structure, but rather a sparser representation containing a fixed percentage with the ``most important'' edges, thus facilitating interpretations. In such cases, consistency is not a criterion that is well-adapted to the problem at hand. Second, we have focused in this paper on graph learning through the graphical Lasso, and the SGL method that imposes further Laplacian constraints. It would be interesting to extend these methods, in order to let the graphical structure depend on well-chosen covariates (e.g., temperature in our river network application, or the time period index in our currency network application), so that the estimated (non-stationary) risk network can then be interpreted through the glasses of these covariates. Moreover, unlike our currency application, where separate model fits were obtained for each time period under consideration, introducing covariates in the graph learning procedure would allow estimating the networks simultaneously from the combined dataset, thus gaining efficiency for higher accuracy of the estimated graph structure. A possibility could be to extend the Gaussian graphical regression approach proposed by \citet{Zhang.Li:2022} to our PTCC-based extremes framework. Third, by analogy with the widely used Gaussian Markov random field framework, one could imagine constructing Markov random fields for extremes where the distribution of connected variables may be fitted with an asymptotically justified model for extremes (e.g., of multivariate Pareto type), whereas the factorization of the likelihood could be pre-specified by the extremal network learned from the data in a preliminary step using our proposed method. Finally, another direction to investigate concerns the geometric representation of multivariate extremes \citep{nolde2020linking, simpson2021geometric}. It would be interesting to see if the new PTCC can be defined through this geometric approach and to explore its links with other popular measures of extremal dependence.

\section*{Acknowledgments}
We point out that there has been independent and parallel work by Lee \& Cooley, who also investigate partial tail correlation for extremes. Our understanding is that inference in their work focuses rather on hypothesis testing (with the goal of checking if the partial tail correlations for given pairs of variables are significantly different from zero), and not on learning extremal networks with structural constraints. This publication is based upon work supported by the King Abdullah University of Science and Technology (KAUST) Office of Sponsored Research (OSR) under Award No. OSR-CRG2020-4394.

\section*{Disclosure statement}
The authors report there are no competing interests to declare.

\bibliographystyle{CUP}
\bibliography{Biblio}


\clearpage
\section*{Appendix}
\appendix
\section{Data details} \label{appd:currency}
\begin{table}[h]
\centering
\begin{tabular}{ccc} 
\hline
Countries & Currency symbol & Currency name\\
\hline
EU & EUR & EURO \\
United Kingdom & GBP & Pound Sterling \\
India & INR & Indian Rupee \\
Australia & AUD & Australian Dollar \\
Canada & CAD & Canadian Dollar \\
South Africa & ZAR & Rand \\
Japan & JPY & Yen \\
Singapore & SGD & Singapore Dollar \\
China & CNY & Yuan\ \\
Switzerland & CHF & Swiss Franc \\
Republic of Korea & KRW & Won \\
Turkey & TRY & Turkish Lira \\
Mexico & MXN & Mexican Peso \\
Brazil & BRL & Real \\
Indonesia & IDR & Rupiah \\
Saudi Arabia & SAR & Saudi Riyal \\
Russia & RUB & Ruble \\
Argentina & ARS & Argentine Peso \\
Ukraine & UAH & Ukrainian Hryvnia \\
Kazakstan & KZT & Kazakhstani Tenge\\
\hline
\end{tabular}
\caption{Currency symbol list.}
\label{table:currency}
\end{table}

\begin{figure}[H]
	\centering
	\includegraphics[width=0.8\linewidth]{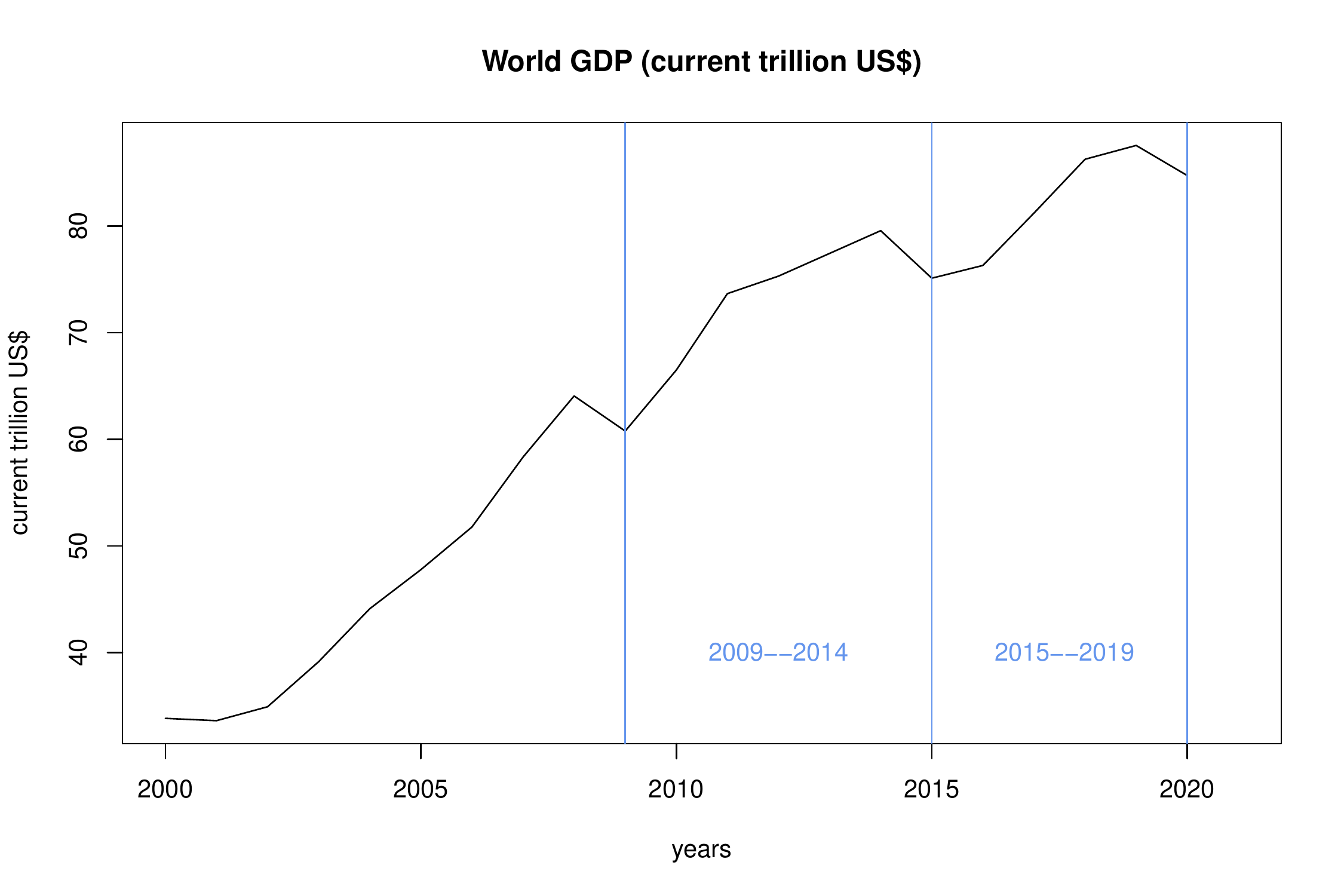}
	\caption{World GDP (current trillion US\$) from 2000 to 2020. Data source: \href{https://data.worldbank.org/indicator/NY.GDP.MKTP.CD}{The World Bank}.}\label{fig:gdp}
\end{figure}

	\end{document}